\documentclass[aps,prd,twocolumn,groupedaddress,nofootinbib]{revtex4-1}

\usepackage[english,american]{babel}

\usepackage{amsmath}
\usepackage{graphicx}
\usepackage[colorlinks=true, allcolors=blue]{hyperref}
\usepackage{verbatim}
\usepackage[T1]{fontenc}
\usepackage[utf8]{inputenc}
\usepackage[american]{babel}
\usepackage{epsfig}
\usepackage{graphicx}
\usepackage{booktabs}
\usepackage{multirow}
\usepackage{dcolumn}
\usepackage{amsmath}
\usepackage{mathtools}
\usepackage{amsfonts}
\usepackage{amssymb}
\usepackage{epstopdf}
\usepackage{bm}
\usepackage{siunitx}
\usepackage{braket}
\usepackage{enumitem}
\usepackage{soul}
\usepackage{booktabs}
\usepackage[table]{xcolor}
\usepackage{color}
\usepackage{transparent}
\usepackage{pifont}
\usepackage{todonotes}
\usepackage[normalem]{ulem}
\usepackage{pbox}
\usepackage{scalerel}
\usepackage{subfiles}

\definecolor{navyblue}{rgb}{0.0, 0.0, 0.5}
\definecolor{royalblue}{rgb}{0.25, 0.41, 0.88}
\definecolor{cadmiumgreen}{rgb}{0.0, 0.42, 0.24}
\definecolor{blue-violet}{rgb}{0.54, 0.17, 0.89}
\definecolor{darkviolet}{rgb}{0.58, 0.0, 0.83}
\definecolor{orange(colorwheel)}{rgb}{1.0, 0.5, 0.0}

\usepackage{hyperref}
\hypersetup{
    colorlinks=true, 
    linkcolor=royalblue, 
    citecolor=magenta}
\usepackage[capitalize]{cleveref}
\usepackage{multirow}

\renewcommand\[{\left[}

\newcommand\ie{{\it i.e.}~}
\newcommand\eg{{\it e.g.}~}

\usepackage{booktabs}
\usepackage{multirow}
\usepackage{dcolumn}
\usepackage{colortbl}

\definecolor{magenta(process)}{rgb}{1.0, 0.0, 0.56}

\definecolor{darkspringgreen}{rgb}{0.09, 0.45, 0.27}

\definecolor{royalblue(web)}{rgb}{0.25, 0.41, 0.88}

\begin{document}

\title{Revisiting constraints on WIMPs around primordial black holes}

\author{Estanis Utrilla Gin\'{e}s}

\affiliation{Instituto de F\'{i}sica Corpuscular (IFIC), University of Valencia-CSIC, Parc Cient\'{i}fic UV, c/ Cate\-dr\'{a}tico Jos\'{e} Beltr\'{a}n 2, E-46980 Paterna, Spain}
\author{Samuel J.~Witte}
\affiliation{Gravitation Astroparticle Physics Amsterdam (GRAPPA), Institute for Theoretical Physics Amsterdam and Delta Institute for Theoretical Physics, University of Amsterdam, Science Park 904, 1098 XH Amsterdam, The Netherlands}
\affiliation{Excellence Cluster ORIGINS, Boltzmannstr. 2, D-85748 Garching, Germany}
\author{Olga Mena}
\affiliation{Instituto de F\'{i}sica Corpuscular (IFIC), University of Valencia-CSIC, Parc Cient\'{i}fic UV, c/ Cate\-dr\'{a}tico Jos\'{e} Beltr\'{a}n 2, E-46980 Paterna, Spain}
\begin{abstract} 
While Primordial Black Holes (PBHs) with masses $M_{\rm PBH} \gtrsim 10^{-11} \, M_\odot$ cannot comprise the entirety of dark matter, the existence of even a small population of these objects can have profound astrophysical consequences. A sub-dominant population of PBHs will efficiently accrete dark matter particles  before matter-radiation equality, giving rise to high-density dark matter spikes. We consider here the scenario in which dark matter is comprised primarily of Weakly Interacting Massive Particles (WIMPs) with a small sub-dominant contribution coming from PBHs, and  revisit the constraints on the annihilation of WIMPs in these spikes using observations of the isotropic gamma-ray background (IGRB) and the Cosmic Microwave Background (CMB), for a range of WIMP masses, annihilation channels, cross sections, and PBH mass functions.
We find that the constraints derived using the IGRB have been significantly overestimated (in some cases by many orders of magnitude), and that limits obtained using observations of the CMB are typically stronger than, or comparable to, those coming from the IGRB. Importantly, we show that $\sim \mathcal{O}(M_\odot)$ PBHs can still contribute significantly to the dark matter density for sufficiently low WIMP masses and p-wave annihilation cross sections.
\end{abstract}
\maketitle

\section{Introduction}

The canonical cosmological model assumes that cold dark matter (CDM) is comprised of a non-relativistic gas of weakly interacting particles. Despite its simplicity, this minimal scenario provides an excellent fit to both CMB and large-scale structure measurements~\cite{Adam:2015rua, Ade:2015xua, Aghanim:2015xee, Alam:2016hwk,Aghanim:2018eyx}. However, a precise understanding of the fundamental nature of dark matter is missing and remains at the forefront in the current list of unsolved problems in modern physics.

Although dark matter is usually interpreted in terms of a new elementary particle, other alternatives exist. Black holes produced from the collapse of large over-densities seeded prior to Big Bang Nucleosynthesis (BBN)\footnote{Several mechanisms have been proposed to generate the initial seed fluctuations, including \eg inflation~\cite{Carr:1993aq, Carr:1994ar, Ivanov:1994pa, Yokoyama:1995ex, GarciaBellido:1996qt, Taruya:1998cz, Green:2000he, Bassett:2000ha}, the collapse of domain walls~\cite{Sato:1981bf, Maeda:1981gw, Berezin:1982ur} and cosmic strings~\cite{Hogan:1984zb, Hawking:1987bn, Polnarev:1988dh}, first-order phase transitions~\cite{Crawford:1982yz, Hawking:1982ga, Kodama:1982sf, Hall:1989hr, Moss:1994iq, Konoplich:1999qq, Jedamzik:1999am, Khlopov:2000js}, and the decays of non-topological solitons~\cite{Cotner:2016cvr,Cotner:2018vug}.}, i.e.\ Primordial Black Holes (PBHs), represent such an alternative --- remarkably, this solution is as old as particle dark matter~\cite{Chapline:1975}. This possibility has recently attracted much attention~\cite{Bird:2016dcv, Clesse:2016vqa, Sasaki:2016jop, Raidal:2017mfl} in the context of the LIGO and VIRGO discoveries of several binary black hole mergers~\cite{Abbott:2016blz, TheLIGOScientific:2016pea, Abbott:2017vtc, Abbott:2016nmj, Abbott:2017oio}. Should the PBHs have masses $\lesssim 10^{-16} M_\odot$, they will efficiently emit Hawking radiation~\cite{Hawking:1974rv, Hawking:1974sw} and evaporate on cosmological timescales -- this process leads to strong energy injection (see \eg~\cite{Khlopov:2008qy,Carr:2009jm,Clark:2016nst,Poulin:2016anj,Boudaud:2018hqb,DeRocco:2019fjq,Laha:2019ssq,Laha:2020ivk}), severely limiting the abundance of PBHs in this regime. Heavier PBHs can imprint observational signatures in a variety of different manners, including via gravitational lensing~\cite{Green:2017qoa, Garcia-Bellido:2017xvr, Zumalacarregui:2017qqd, Garcia-Bellido:2017imq}, the dynamical evolution of gravitionatially bound systems~\cite{Monroy-Rodriguez:2014ula, Brandt:2016aco, Green:2016xgy, Li:2016utv, Koushiappas:2017chw, Kavanagh:2018ggo}, observable radio and x-ray emission~\cite{Gaggero:2016dpq, Inoue:2017csr, Hektor:2018rul}, and spectral distortions in the CMB~\cite{Tada:2015noa, Young:2015kda, Chen:2016pud, Ali-Haimoud:2016mbv, Blum:2016cjs, Horowitz:2016lib, Poulin:2017bwe, Bernal:2017vvn, Nakama:2017xvq, Deng:2018cxb}. Collectively, these observations prohibit PBHs from constituting the entirety of dark matter, unless their masses are confined roughly to the range $10^{-16} \, M_\odot \lesssim M_{\rm PBH} \lesssim 10^{-11} \, M_\odot$
(see, e.g., Refs~\cite{Belotsky:2014kca, Carr:2016drx, Sasaki:2018dmp, Green:2020jor,Mack:2006gz, Ricotti:2007au, Josan:2009qn, Carr:2009jm, Capela:2013yf, Clesse:2016vqa,  Green:2016xgy, Bellomo:2017zsr, Kuhnel:2017pwq, Carr:2017jsz, Sasaki:2018dmp,Villanueva-Domingo:2021spv} for recent reviews on PBHs).

Despite stringent constraints on the abundance of heavy PBHs, even a small number of these objects can have a significant impact in astrophysics and cosmology. In particular, it has been shown that PBHs can efficiently accrete the primary component of dark matter prior to matter-radiation equality, generating dense dark matter spikes referred to as ultra-compact mini-halos (UCMHs). If dark matter is mostly comprised of Weakly Interacting Massive Particles (WIMPs), the large densities found in the UCMHs will dramatically enhance the efficiency of WIMP annihilation, imprinting powerful observational signatures \eg in gamma-ray flux~\cite{Lacki:2010zf, Josan:2010vn, Boucenna:2017ghj, Eroshenko:2016yve, Carr:2020mqm, Hertzberg:2020kpm, Adamek:2019gns, Kadota:2021jhg} and the anisotropies of the CMB~\cite{Ricotti:2007au, Tashiro:2021xnj}.

In this work we revisit the cosmological and astrophysical constraints on the mixed WIMP-PBH dark matter scenario, focusing in particular on those derived using observations of the extragalactic gamma ray background and the CMB. We incorporate the state-of-the-art understanding of the UCMH density profiles, investigating a wide array of WIMP dark matter models (spanning MeV-scale to TeV-scale WIMP masses, a variety of final states, and both s-wave and p-wave annihilation), for both monochromatic and extended PBH mass functions. Our calculations show that the strength of constraints derived using the extragalactic gamma ray background have been largely overestimated in previous studies in the literature (in some cases by many orders of magnitude), and are typically comparable or sub-dominant to those obtained using the latest observations of the CMB.

This manuscript is organized as follows. Section~\ref{sec:dm_all} outlines the details of WIMP dark matter annihilation in UCHMs around PBHs. Section~\ref{sec:methods} describes the methodology and procedure used to derive constraints on the abundance of PBHs using both the CMB and $\gamma$-ray observations. Section~\ref{sec:results} presents our results, as well as a critical comparison to previous analyses. We conclude in Sec.~\ref{sec:conc}.

\section{Energy Injection from WIMP Annihilation near PBHs}\label{sec:dm_all}

\subsection{WIMP ANNIHILATION}
\label{sec:dm}
Among the most studied and theoretically appealing dark matter candidates are electroweak scale WIMPs, as these particles can be efficiently produced with the correct relic abundance via the thermal freeze-out mechanism and naturally appear in a plethora of well-motivated extensions of the Standard Model~\cite{Kolb:1990vq, Bertone:2010zza}.

The annihilation rate of a Majorana dark matter candidate $\chi$ is given by~\footnote{An additional factor of 1/2 must be included for Dirac dark matter.}
\begin{equation} \label{ann_rate}
\Gamma_{\mathrm{ann}} \equiv \frac{1}{2 m_\chi^2} \int_V dV \left\langle \sigma_A v \right\rangle \rho_\chi^2~,
\end{equation}
where $m_\chi$ and $\rho_\chi$ are the WIMP mass and density, and $\left\langle \sigma_A v \right\rangle$ is the thermally averaged annihilation cross section. At freeze-out, the annihilation cross section is given by $\left\langle \sigma_A v \right\rangle \simeq 3\times 10^{-26}/f_\chi \; [\mathrm{cm}^3 \mathrm{s}^{-1}]$ where $f_\chi$ is the fraction of dark matter in the form of WIMPs. The energy density injected per unit time into a species $c$ in the energy range $[E_1, E_2]$ from WIMPs annihilating in UCMHs is given by
\begin{equation} \label{eq:dE_dV}
\frac{dE_{\scaleto{\mathrm{UCMH}}{4pt}}}{dVdt}= \int_{E_1}^{E_2} \, dE \, n_{\scaleto{\rm UCMH}{4pt}}\sum_c B_c \, \Gamma_{\rm ann}^{(c)} \frac{dN^{(c)}}{dE} \, .
\end{equation}
Here, $n_{\scaleto{\rm UCMH}{4pt}}$ is the number density of UCMHs, $B_c$ is the branching fraction of WIMP annihilation to species $c$ (e.g. photons, electrons, etc), $\Gamma_{\rm ann}^{(c)}$ is the WIMP annihilation rate in a single UCMH, and $dN^{(c)}/dE$ is the spectra of the final state species. Should the annihilation be s-wave, \ie velocity independent, the annihilation cross section today is given by that at freeze-out. For p-wave annihilating dark matter, we estimate the radially-dependent annihilation cross section by assuming that the local distribution is approximately described by a Maxwell-Boltzmann distribution with a dispersion obtained by applying the virial theorem. We further assume that the gravitational potential is entirely dominated by the PBH contribution at the center of the UCMH (an assumption which we have verified has a negligible effect on the predicted annihilation rate). This allows us to parameterize the thermally averaged p-wave annihilation cross section as
\begin{equation} \label{eq:sigmav_pwave}
\left\langle \sigma_A v \right\rangle^{\mathrm{p-wave}} = \left\langle \sigma_A v \right\rangle_{\mathrm{fo}}\frac{v^2}{v_{\mathrm{fo}}^2}=\frac{\left\langle \sigma_A v \right\rangle_{\mathrm{fo}}}{v_{\mathrm{fo}}^2}\frac{G M_{\scaleto{\mathrm{PBH}}{4pt}} }{r}~,
\end{equation}
 where the velocity dispersion at freeze out is estimated as $v_{\mathrm{fo}} \sim 0.3$~\cite{Kadota:2021jhg}. 

For WIMP masses $\gtrsim 5$~GeV, we compute the annihilation spectra $\frac{dN^{(c)}}{dE}$ in Eq.~(\ref{eq:dE_dV}) above using the publicly available tool PPPC4DMID \cite{Cirelli:2010xx}, and focus for simplicity on the case of pure annihilations into the $b\bar{b}$ and the $e^+e^-$ channels.  In order to explore the parameter space of lighter dark matter candidates at sub-GeV scales we have used the spectra derived with the tool Hazma \cite{Coogan:2019qpu}, which employs chiral perturbation theory (with a Lagrangian that includes the SM, the dark sector, and hadrons) in order to calculate cross sections and spectra from decays, annihilations, and scatterings at next-to-leading order (NLO). The tool assumes the dark matter particle to be a Dirac fermion and considers models of scalar and vector mediators. We focus here on the case of a 100 MeV dark matter particle interacting with the Standard Model through a massive kinetically-mixed vector mediator. The dominant annihilation channel for this candidate is to $e^+ e^-$, but there is also a $<\mathcal{O}(10^{-6})$ suppressed annihilation into the $\pi^0 \gamma$ channel.  The $e^+ e^-$ channel produces a monochromatic $e^\pm$ line at 100 MeV and a continuous gamma ray spectrum of final state radiation/internal bremsstrahlung (FSR/IB). The highly suppressed $\pi^0 \gamma$ channel, on the other hand, contributes negligibly to the photon spectrum. For the this light dark matter candidate we adopt an annihilation cross-section of $\left\langle \sigma_A v \right\rangle=10^{-28}/f_\chi \; [\text{cm}^3 \text{s}^{-1}]$\footnote{It is worth noting that while s-wave annihilation cross sections below the thermal value tend to over produce dark matter, this is not always the case, and can easily be avoided in non-minimal models. }, since this is approximately the maximally allowed value by $\textit{Planck}$~\cite{Planck:2018vyg} (note that p-wave annihilating dark matter is not constrained to this level, however we fix the cross section to have the same value for comparison purposes).  

\subsection{Ultra Compact Mini-Halos around PBHs}
\label{sec:mass}

The total amount of energy produced from dark matter annihilations in UCMHs depends both on the density profile of the UCMH and on the mass distribution of PBHs. We shall discuss each of these below.

PBHs are formed in the early Universe when the cosmological horizon crosses a large enough over-density. Before kinetic decoupling at $t_{\scaleto{\mathrm{KD}}{4pt}}$, the radiation pressure does not allow PBHs to accrete a significant amount of mass~\cite{Eroshenko:2016yve}. However, after $t_{\scaleto{\mathrm{KD}}{4pt}}$, a WIMP spike forms as spherical shells enter the expanding region of influence of the PBH -- this process allows the spike to accrete a total mass that exceeds the PBH mass by up to two orders of magnitude. The region over which the PBH exerts its gravitational influence is approximately defined by the radius $r_{\mathrm{infl}}$ at which WIMPs decouple from the Hubble expansion. We follow Ref.~\cite{Adamek:2019gns} in numerically estimating this radius as $r_{\mathrm{infl}} \simeq (2GM_{\scaleto{\mathrm{PBH}}{4pt}} t^2)^{\frac{1}{3}}~$. At matter-radiation equality the sphere of this radius, $r_{\mathrm{infl}}(t_{\mathrm{eq}}) \simeq (2GM_{\scaleto{\mathrm{PBH}}{4pt}}t_{eq}^2)^{1/3}$, contains a mass comparable to the PBH mass. 

If the kinetic energy of WIMPs is negligible compared to the gravitational potential energy, a power-law density profile scaling like $\rho \sim r^{-9/4}$ develops from the accreting material. N-body simulations and analytical calculations have shown that the $9/4$ density profile is a good approximation in the regime of large PBH and WIMP particle masses~\cite{Adamek:2019gns}. However when the ratio of kinetic to potential energy cannot be neglected, one must consider that particles in the high energy tail may escape the gravitational pull of the PBH, while those at lower energies fall into bound orbits with varying angular momentum. 
These effects were first accounted for in Ref.~\cite{Eroshenko:2016yve} by adopting a Maxwell-Boltzmamn distribution, and integrating the phase-space of bound trajectories over their orbits.
A semi-analytic calculation of the phase-space integral was provided in Ref.~\cite{Boucenna:2017ghj}, and later improved by Refs.~\cite{Boudaud:2021irr, Carr:2020mqm}.
In this work, we shall use the later analytic result, which in general gives rise to a broken triple power-law profile, although dark matter annihilations deplete dark matter (thus setting an upper limit on the dark matter density), such that the final density profile follows a truncated single, double, or triple power law. Importantly, it is the ratio of the kinetic to potential energy which determines whether one expects a single power law profile (with a slope of $9/4$, occurring when kinetic energy is negligibly small), a double broken power law profile (with an inner slope of $3/2$, and an outer slope of $9/4$), or a triple broken power law (with slopes of $3/4$, $3/2$, and $9/4$, appearing at increasing radii, and occurring when the kinetic energy is large relative to the gravitational potential energy).

The analytical expression for the density profile from Refs.~\cite{Boudaud:2021irr, Carr:2020mqm} is given by
\begin{equation} \label{rho_chi}
\begin{split}
\rho_\chi(r) = \left\{
\begin{array}{ll}
f_\chi \rho_{\scaleto{\mathrm{KD}}{4pt}}\left( \frac{r_C}{r }\right)^{\frac{3}{4}}      & \mathrm{for\ } r \le r_{\scaleto{C}{4pt}}~,\\
f_\chi \frac{\rho_{\scaleto{\mathrm{eq}}{4pt}}}{2}\left( \frac{M}{M_\odot}\right)^{\frac{3}{2}}\left( \frac{\hat{r}}{r}\right)^{\frac{3}{2}} & \mathrm{for\ } r_{\scaleto{C}{4pt}} \le r \le r_{\scaleto{K}{4pt}}~,\\
f_\chi \frac{\rho_{\scaleto{\mathrm{eq}}{4pt}}}{2}\left( \frac{M}{M_\odot}\right)^{\frac{3}{4}}\left( \frac{\bar{r}}{r}\right)^{\frac{9}{4}}     & \mathrm{for\ } r > r_{\scaleto{K}{4pt}}~,
\end{array}
\right.
\end{split}
\end{equation}\\
 with $\hat{r}$ and $\bar{r}$ defined as
\begin{equation} \label{r_hat}
\hat{r} \equiv G M_\odot \frac{t_{\scaleto{\mathrm{eq}}{4pt}}}{t_{\scaleto{\mathrm{KD}}{4pt}}} \frac{m_\chi}{T_{\scaleto{\mathrm{KD}}{4pt}}}~, \;\;\;\; \bar{r} \equiv (2GM_\odot t^2_{\scaleto{\mathrm{eq}}{4pt}})^{\frac{1}{3}}~.
\end{equation}
Equating the first two analytic profiles of Eq.~(\ref{rho_chi}) and the last two, it is possible to obtain the values for $r_{\scaleto{C}{4pt}}$ and $r_{\scaleto{K}{4pt}}$, which are given by
\begin{equation} \label{r_C}
r_{\scaleto{C}{4pt}} = \frac{r_{\scaleto{S}{4pt}}}{2}\left(\frac {m_\chi}{T_{\scaleto{\mathrm{KD}}{4pt}}}\right)~, \;\;\;\; r_{\scaleto{K}{4pt}} = 4\frac{t_{\scaleto{\mathrm{KD}}{4pt}}^2 }{r_{\scaleto{S}{4pt}} }\left(\frac{T_{\scaleto{\mathrm{KD}}{4pt}}}{m_\chi}\right)^2 \, ,
\end{equation}
where $r_s$ is the the Schwartschild radius. The time $t_{\scaleto{\mathrm{KD}}{4pt}}$  and temperature $T_{\scaleto{\mathrm{KD}}{4pt}}$ at kinetic decoupling are approximately given by  \cite{Boucenna:2017ghj} 
\begin{equation} \label{TKD}
T_{\scaleto{\mathrm{KD}}{4pt}} = \frac{m_\chi}{\Gamma[3/4]} \left(\frac{\alpha \cdot m_\chi}{M_{\scaleto{\mathrm{Pl}}{4pt}}}\right)^{\frac{1}{4}}, \;\;\;\; t_{\scaleto{\mathrm{KD}}{4pt}} = \frac{2.4}{\sqrt{g_{\scaleto{\mathrm{KD}}{4pt}}}} \left(\frac{T_{\scaleto{\mathrm{KD}}{4pt}}}{1 \; \mathrm{MeV}}\right)^{-2}~,
\end{equation}\\
 with $\alpha = (16\pi^3 \, g_{\scaleto{\mathrm{KD}}{4pt}}/45)^{1/2}$. For the sake of simplicity, we have fixed the relativistic degrees of freedom at kinetic decoupling to be $g_{\scaleto{\mathrm{KD}}{4pt}} = 61.75$.

\begin{figure*}
	\centering
	   \includegraphics[width=0.44\textwidth]{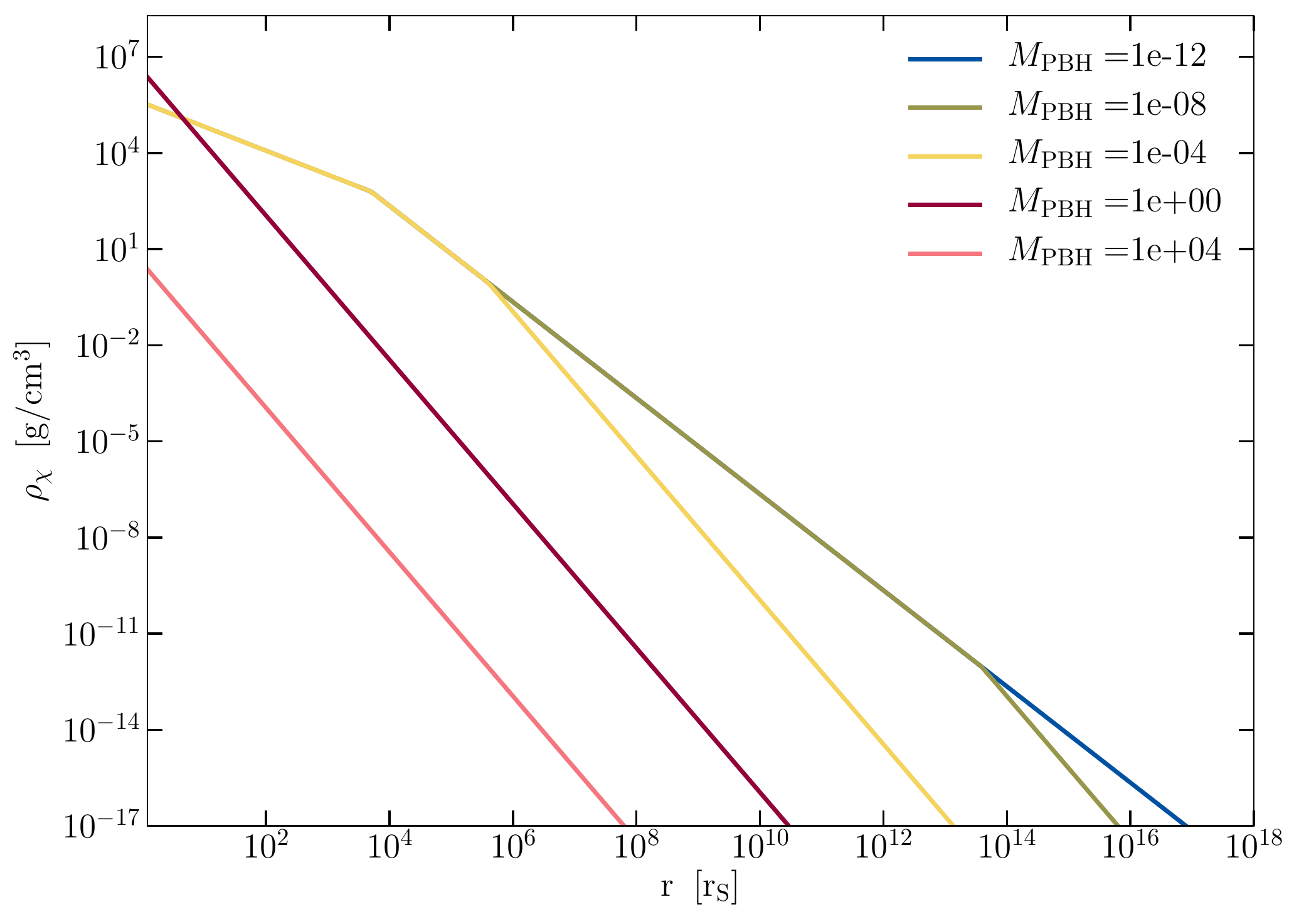}
	   \includegraphics[width=0.44\textwidth]{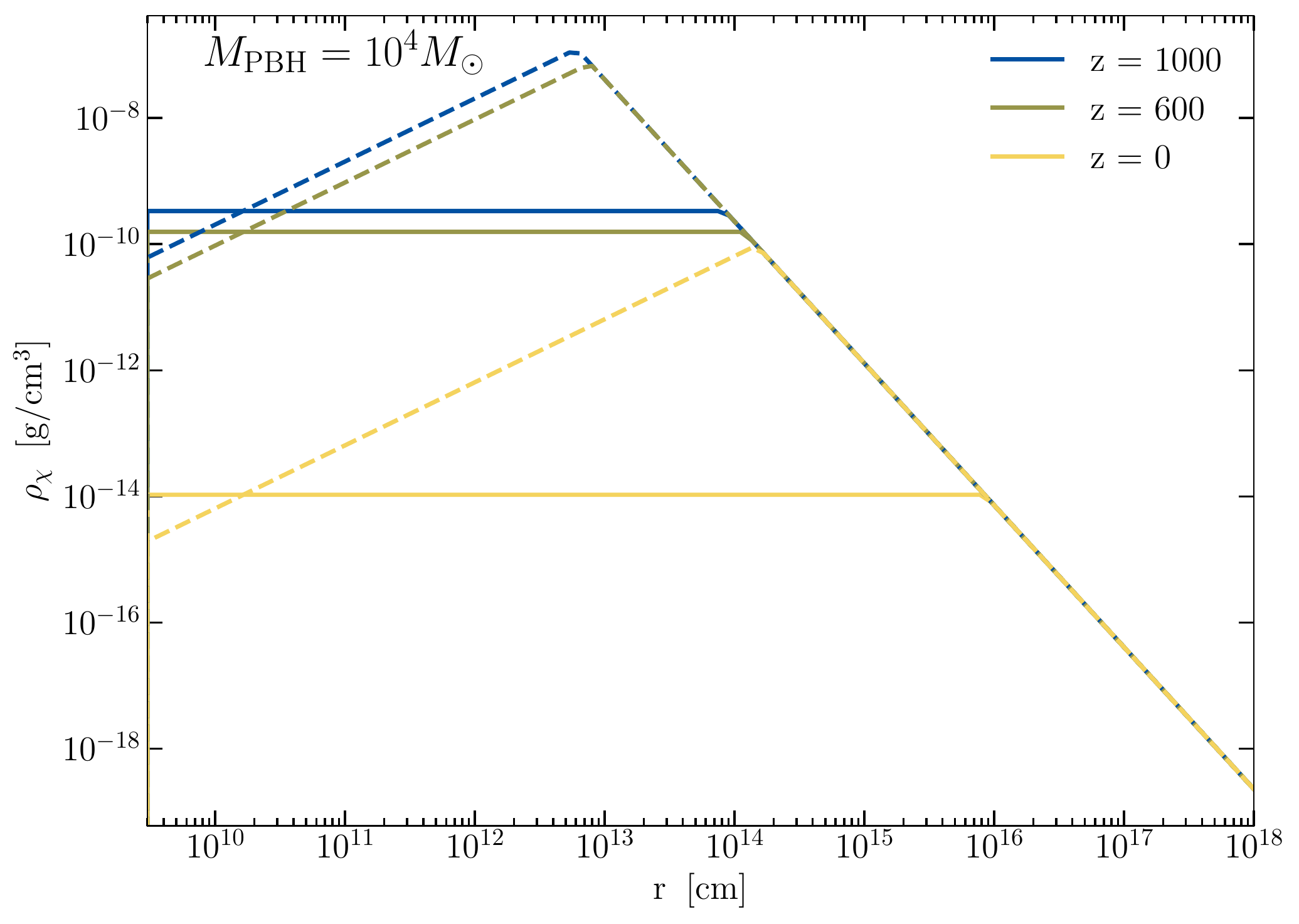}
	\caption{The left (right) panel depicts the density profile before (after) WIMP annihilation as a function of radial distance. Results are illustrated  for several PBH masses (left), and several redshifts for $M_{\rm PBH} = 10^4 \, M_\odot$ (right). In the right panel solid and dashed lines denote the effect of s-wave and p-wave WIMP annihilations, respectively.}
	\label{fig:rho}
\end{figure*}

The resulting density profiles are depicted in Fig.~\ref{fig:rho}. In the left panel we show the density as a function of radius (in units of $r_s$) for various PBH masses, while in the right panel we illustrate how WIMP annihilations modify the density profile in the central region of the mini-halo. This effect is shown for a $10^4 \, M_\odot$ PBH at various redshifts, and assuming both s-wave (solid lines) and p-wave (dashed lines) annihilating WIMP dark matter. The presence of annihilations saturates the density to a maximum value $\rho_{\mathrm{max}}$, which is roughly given by \cite{Bertone:2005xz}
\begin{equation} \label{rho_max}
\rho_{\mathrm{max}} = \frac{m_\chi}{\left\langle \sigma_A v \right\rangle t} \, ,
\end{equation}
 where $t$ is the age of the PBH (which we approximate here to be the age of the Universe). We note that the annihilation rate from an individual UCMH is dominated by the largest radius which for which the density profile is saturated to $\rho_{\mathrm{max}}$.
 
Notice, from the right panel of Fig.~\ref{fig:rho}, that the p-wave annihilation channel has two main effects on the annihilation rate. Firstly, the density profile near the PBH grows radially (unlike in the case of s-wave annihilations, where it is flat) due to the velocity dependence of the annihilation cross section, allowing the p-wave profile to reach larger densities than in the case of s-wave annihilations. The enhancement in the annihilation rate from the larger densities, however, is offset by the suppression of the velocity averaged annihilation cross section.

The other crucial ingredient when modeling the net energy injection is the mass distribution of the PBHs. While a monochromatic PBH mass function is the most commonly adopted distribution, this is an unphysical choice motivated only for simplicity. In this work we adopt both a monochromatic mass distribution, used for sake of comparison with the broader literature on PBHs, and the more physically motivated log-normal mass function given by~\cite{Carr:2017jsz}:
\begin{equation} \label{psi}
\begin{split}
\psi(M) \equiv \frac{1}{\bar{\rho}_{\scaleto{\mathrm{PBH}}{4pt}}} \frac{d\rho(M)}{dM}= \frac{1}{\sqrt{2\pi}\sigma M} \mathrm{Exp}\left({-\frac{\mathrm{Log}^2(M/M_{\mathrm{pk}})}{2\sigma^2}}\right)~,
\end{split}
\end{equation}
\noindent where $M_{\mathrm{pk}}$ and $\sigma$ are the mass at the peak of the spectrum and its width, respectively. The analytic derivation of the s-wave annihilating rates in the monochromatic and broad PBH mass function cases are detailed in Appendix \ref{sec:apb} and \ref{sec:apd}, respectively (analytic expressions for p-wave annihilating dark matter and a monochromatic PBH mass function is also provided in Appendix \ref{sec:apc}).

\section{Methodology}\label{sec:methods}

\subsection{Constraints from CMB}

Since the energy deposition from the WIMP annihilation products occurs on scales much longer than the inter-UCMH distance, one can treat the cumulative energy injection from all UCMHs as uniform in the Inter Galactic Medium (IGM). One can express the rate of energy deposition per unit volume as
\begin{equation} \label{deposited}
\left.\frac{dE}{dVdt}\right|_{\mathrm{dep}}(z) = \left.f(z)\frac{dE}{dVdt}\right\rvert_{\mathrm{inj}}(z)~,
\end{equation}
with $f(z)$, the energy deposition function, given by \cite{Slatyer:2012yq}
\begin{equation} \label{deposited_func}
\begin{split}
f(z) = \frac{\int d ln(1+z^{\prime})\frac{(1+z^{\prime})^3}{H(z^{\prime})}(1+B(z^{\prime}))}{\frac{(1+z)^3}{H(z)}(1+B(z))}  \\
\, \times \, \frac{\sum_{l}\int T^{(l)}(z^{\prime}, z, E)E \left.\frac{dN}{dE}\right|^{(l)}_{\mathrm{inj}}dE}{\sum_l \int \left.E\frac{dN}{dE}\right|^{(l)}_{\mathrm{inj}}} \, .
\end{split}
\end{equation}
 Here, $B$ is the boost factor and $dN/dE$ the energy spectrum of the different annihilation products, and we have introduced the transfer functions $T^{(l)}(z^{\prime}, z, E)$. The index $l$ identifies the photon and electron/positron final states. In the following, we use the publicly available transfer functions tabulated by Ref.~\cite{Slatyer:2015kla} to compute the fraction of the deposited energy going into heating, Lyman-$\alpha$ excitation, ionization of the neutral hydrogen and the ionization of the neutral helium. The boost factor, which governs the additional  energy injection arising from annihilations in UCMHs (with respect to the isotropic background), is given by
\begin{equation} \label{e7}
\begin{split}
B &\equiv \left.\frac{dE}{dVdt}\right\rvert_{\mathrm{inj}} \left(\left.\frac{dE}{dVdt}\right\rvert_{\mathrm{bkg}}\right)^{-1} \\ &= \frac{\Gamma_{\mathrm{ann}} f_{\scaleto{\mathrm{PBH}}{4pt}} \rho_{\scaleto{\mathrm{DM,0}}{5pt}} (1+z)^3}{ M_{\scaleto{\mathrm{PBH}}{4pt}}} \left( \left\langle \sigma_A v \right\rangle \frac{\rho_{\scaleto{\mathrm{DM,0}}{5pt}}^2 (1+z)^6}{2 m_\chi^2} \right)^{-1} \, .
\end{split}
\end{equation}
We define $f_{\scaleto{\mathrm{PBH}}{4pt}} \equiv \Omega_{\scaleto{\mathrm{PBH}}{4pt}}/\Omega_{\scaleto{\mathrm{DM}}{4pt}}$ as the  redshift independent PBH fraction, and since we are considering two possible dark matter contributions, $f_{\scaleto{\mathrm{PBH}}{4pt}}= 1-f_\chi$. In the equation above, $\rho_{\scaleto{\mathrm{DM,0}}{5pt}}$ refers to the current dark (total) matter density and $M_{\scaleto{\mathrm{PBH}}{4pt}}$ is the mass (mean mass) of the PBH for the case of a monochromatic (broad) PBH mass function. 

In order to derive the limits from cosmological observations, we have modified the Boltzmann code CLASS~\cite{Blas:2011rf} using its ExoCLASS package~\cite{Stocker:2018avm} to include the calculation of the redshift dependent energy deposition functions, annihilation rates, and boost factors of the different models analyzed in this work. ExoCLASS relies on the recombination code RECFAST~\cite{Seager:1999bc}, which traces the cosmological evolution of the free electron fraction and gas temperature. 
We perform Markov Chain Monte Carlo (MCMCs) likelihood analyses using the publicly available package MontePython~\cite{Audren:2012wb}. 

We vary eight parameters: six from the canonical $\Lambda \rm CDM$ model ($\Omega_b$, $\Omega_{\scaleto{\mathrm{CDM}}{4pt}}$, $H_0$, $\log(10^{10}A_S)$, $n_S$ and $\tau_{\scaleto{\mathrm{reio}}{4pt}}$) plus two model-dependent parameters, $f_{\scaleto{\mathrm{PBH}}{4pt}}$ and $M_{\scaleto{\mathrm{PBH}}{4pt}}$. In addition, we have two parameters which are implicitly derived from $f_{\scaleto{\mathrm{PBH}}{4pt}}$. These are $f_\chi$ and $\left\langle \sigma_A v \right\rangle$. In our analysis we use the following data sets:
\begin{itemize}
\item The Cosmic Microwave Background (CMB) temperature and polarization power spectra from the final release of $\textit{Planck}$ 2018 (in particular we adopt the plikTTTEEE+lowl+lowE likelihood) \cite{Aghanim:2018eyx,Aghanim:2019ame}, plus the CMB lensing reconstruction from the four-point correlation function~\cite{Aghanim:2018oex}.
\item Baryon Acoustic Oscillations (BAO) distance and expansion rate measurements from the 6dFGS~\cite{Beutler:2011hx}, SDSS-DR7 MGS~\cite{Ross:2014qpa}, BOSS DR12~\cite{Alam:2016hwk} galaxy surveys,
as well as from the eBOSS DR14 Lyman-$\alpha$ (Ly$\alpha$) absorption~\cite{Agathe:2019vsu} and Ly$\alpha$-quasars cross-correlation~\cite{Blomqvist:2019rah}. These consist of isotropic BAO measurements of $D_V(z)/r_d$
(with $D_V(z)$ and $r_d$ the spherically averaged volume distance and sound horizon at baryon drag, respectively)
for 6dFGS and MGS, and anisotropic BAO measurements of $D_M(z)/r_d$ and $D_H(z)/r_d$
(with $D_M(z)$ the comoving angular diameter distance and $D_H(z)=c/H(z)$ the radial distance)
for BOSS DR12, eBOSS DR14 Ly$\alpha$, and eBOSS DR14 Ly$\alpha$-quasars cross-correlation. 
\end{itemize}

\begin{figure}
\centering
	\includegraphics[width=\linewidth]{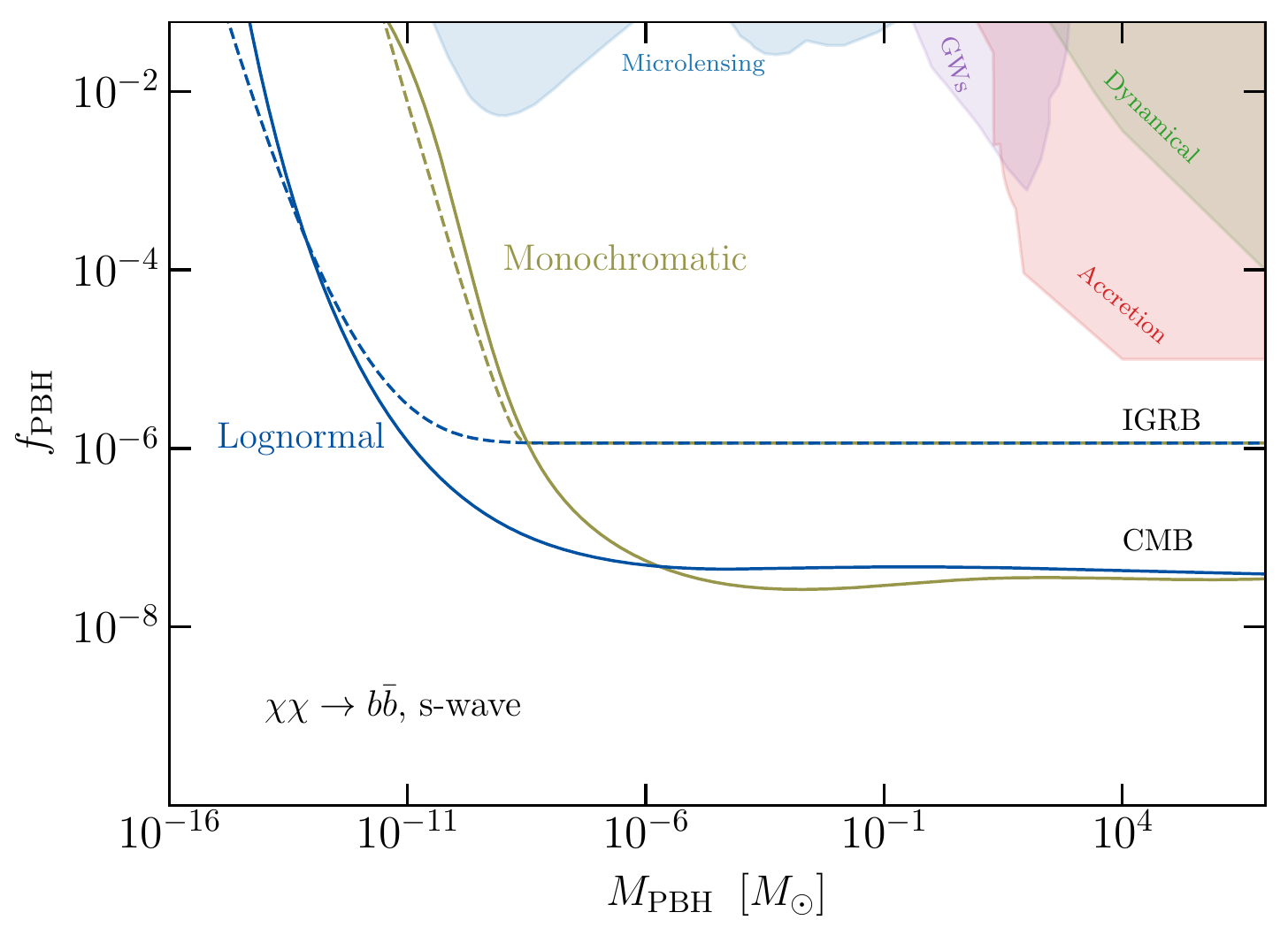}
	\caption{$95\%$~CL constraints on the fraction of PBHs in the form of dark matter $f_{\rm PBH}$ as a function of PBH mass, assuming a monochromatic and log-normal (with $\sigma = 2$) mass functions (shown in green and blue solid lines, respectively). Results are shown assuming the remaining dark matter is made of a $100$~GeV WIMP with s-wave annihilations to $b\bar{b}$. Results are compared to the limits one would obtain using a maximally conservative treatment of the isotropic gamma-ray background (dashed), see Sec.~\ref{sec:3b} for a more optimistic analysis.  We also illustrate the constraints  on the fraction of PBHs in the form of dark matter in the absence of UCMHs from a number of observational probes, see Ref.~\cite{Green:2020jor}. }
	\label{fig:results_broad}
\end{figure}

\subsection{Constraints from $\gamma$-ray observations}
\label{sec:3b}
A highly complementary probe of WIMP annihilation in UCMHs comes from the isotropic $\gamma$-ray background (IGRB), which has historically been the main observation used to constrain the hybrid WIMP-PBH dark matter scenario (see \eg\cite{Lacki:2010zf, Josan:2010vn, Boucenna:2017ghj, Eroshenko:2016yve, Carr:2020mqm, Hertzberg:2020kpm, Adamek:2019gns, Kadota:2021jhg}). In order to highlight the benefits and drawbacks of the cosmological analysis presented here, we re-derive these IGRB constraints, showing that previous analyses have significantly over-estimated the sensitivity.

The IGRB is obtained by removing all the extragalactic resolved point sources from the extragalactic $\gamma$-ray background (EGRB). In this work we make use of the 50-month \textit{Fermi} Large Area Telescope (LAT) IGRB measurements, which have been obtained using the galactic diffuse emission model A from Ref.~\cite{Fermi-LAT:2015qzw} for the full energy range (spanning from 100 MeV to 820 GeV).

We derive constraints using two approaches. Firstly, we adopt a maximally conservative approach, in which we  make no further assumptions about the unresolved astrophysical contribution. Secondly, a more optimistic approach in which the contribution from unresolved extragalactic sources is subtracted from the EGRB is also considered. In the following, we shall refer to the more conservative model just as IGRB (using the \textit{Fermi}-LAT terminology) and to the one with a background model as \emph{optimistic} IGRB.

In our conservative approach, we require the integrated flux in every bin, as computed in Appendix~\ref{sec:ape}, not exceed the observed flux from the IGRB \textit{Fermi}-LAT~\cite{Fermi-LAT:2014ryh}. We do not attempt at this point to account for correlations among bins, but rather simply apply the $2\sigma$ upper limits in each bin, which account for both systematic and statistical uncertainties~\footnote{Example spectra are shown for a few cases alongside the observed data in Appendix~\ref{sec:ape}.} 

Our \emph{optimistic} approach follows the procedure outlined in Refs.~\cite{Boucenna:2017ghj,Adamek:2019gns,Kadota:2021jhg}. The idea here is to map the constraints on decaying dark matter obtained using \textit{Fermi}-LAT observations in the conservative IGRB model into constraints of WIMP annihilation near PBHs. In particular, Refs.~\cite{Boucenna:2017ghj,Adamek:2019gns,Kadota:2021jhg} derive upper limits for the PBH fraction from the results of  Ref.~\cite{Ando:2015qda}, which derives lower limits on the lifetime of decaying dark matter. In order to apply these results to the case of WIMP annihilations, one must assume the annihilation rate is redshift independent; this assumption allows one to relate the constraints on the decay rate to those on $f_{\scaleto{\mathrm{PBH}}{4pt}}$ via

\begin{equation} \label{eq:decay_ann}
f_{\scaleto{\mathrm{PBH}}{4pt}}=\frac{\Gamma_{\mathrm{dec}}M_{\scaleto{\mathrm{PBH}}{4pt}}}{\Gamma_{\mathrm{ann}}m_ \chi} \, .
\end{equation}
In the high PBH mass regime the annihilation rate scales approximately as $(1+z)$, and therefore the assumption that  the annihilation rate is redshift independent is not strictly valid. The error introduced using this procedure yields a result larger by a factor of $\sim 1.5$. We correct for this in what follows by re-scaling the inferred constraint on $f_{\rm PBH}$ after applying Eq.~(\ref{eq:decay_ann}). 

\section{Results and comparison with previous analyses}\label{sec:results}

\begin{figure*}
	\centering
		\includegraphics[width=0.48\linewidth]{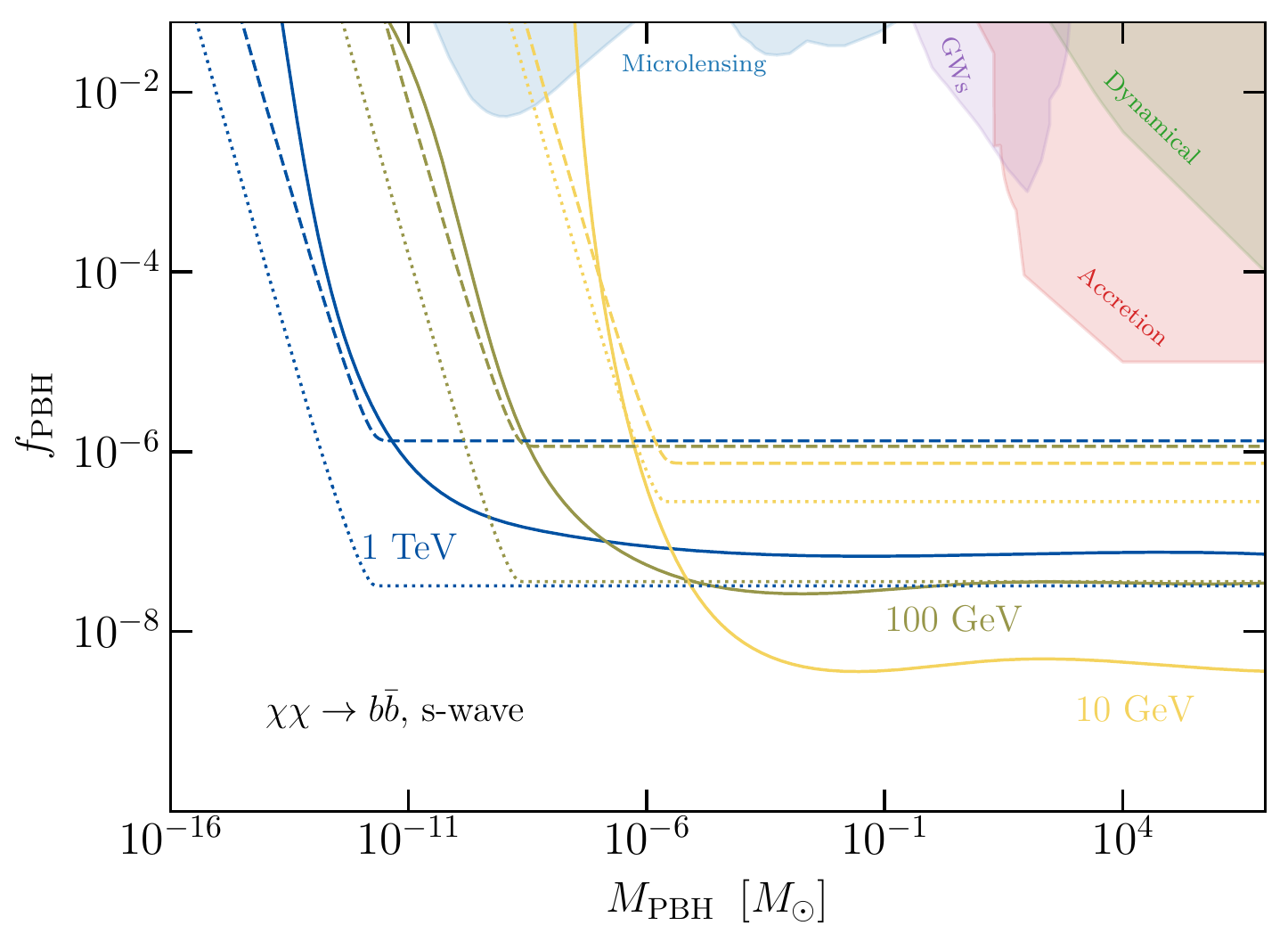}	
		\includegraphics[width=0.48\linewidth]{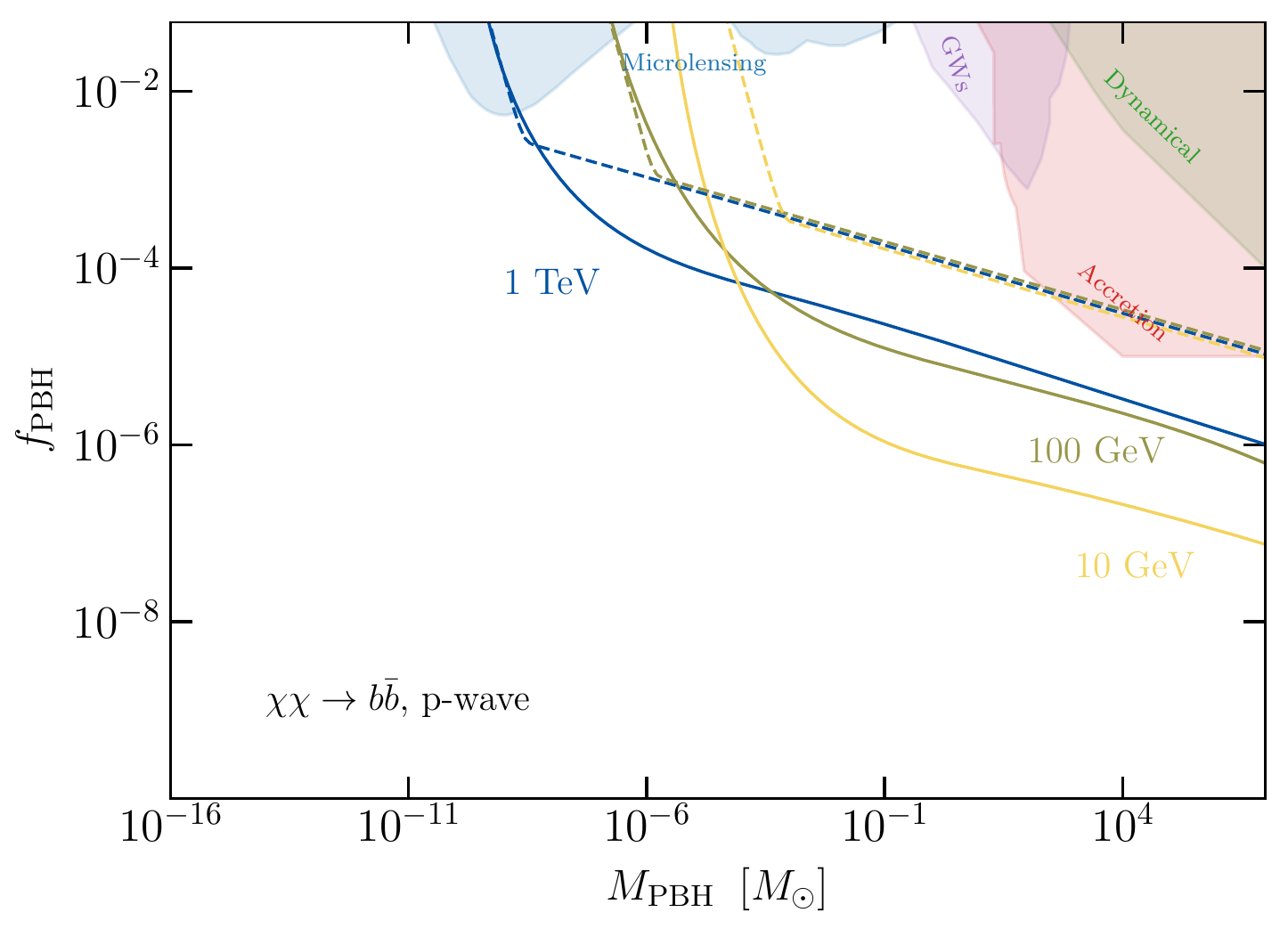}
		\includegraphics[width=0.48\linewidth]{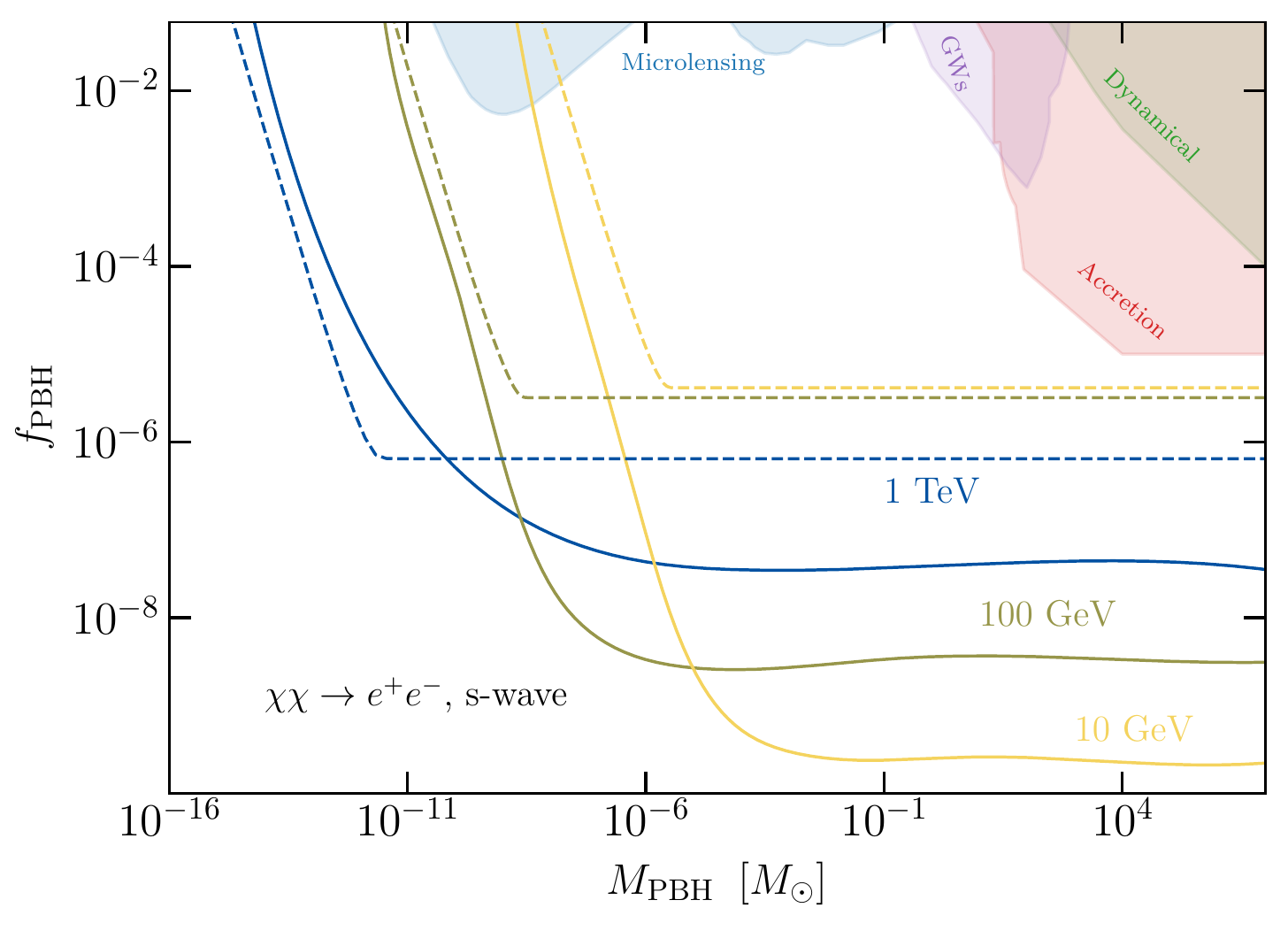}	
		\includegraphics[width=0.48\linewidth]{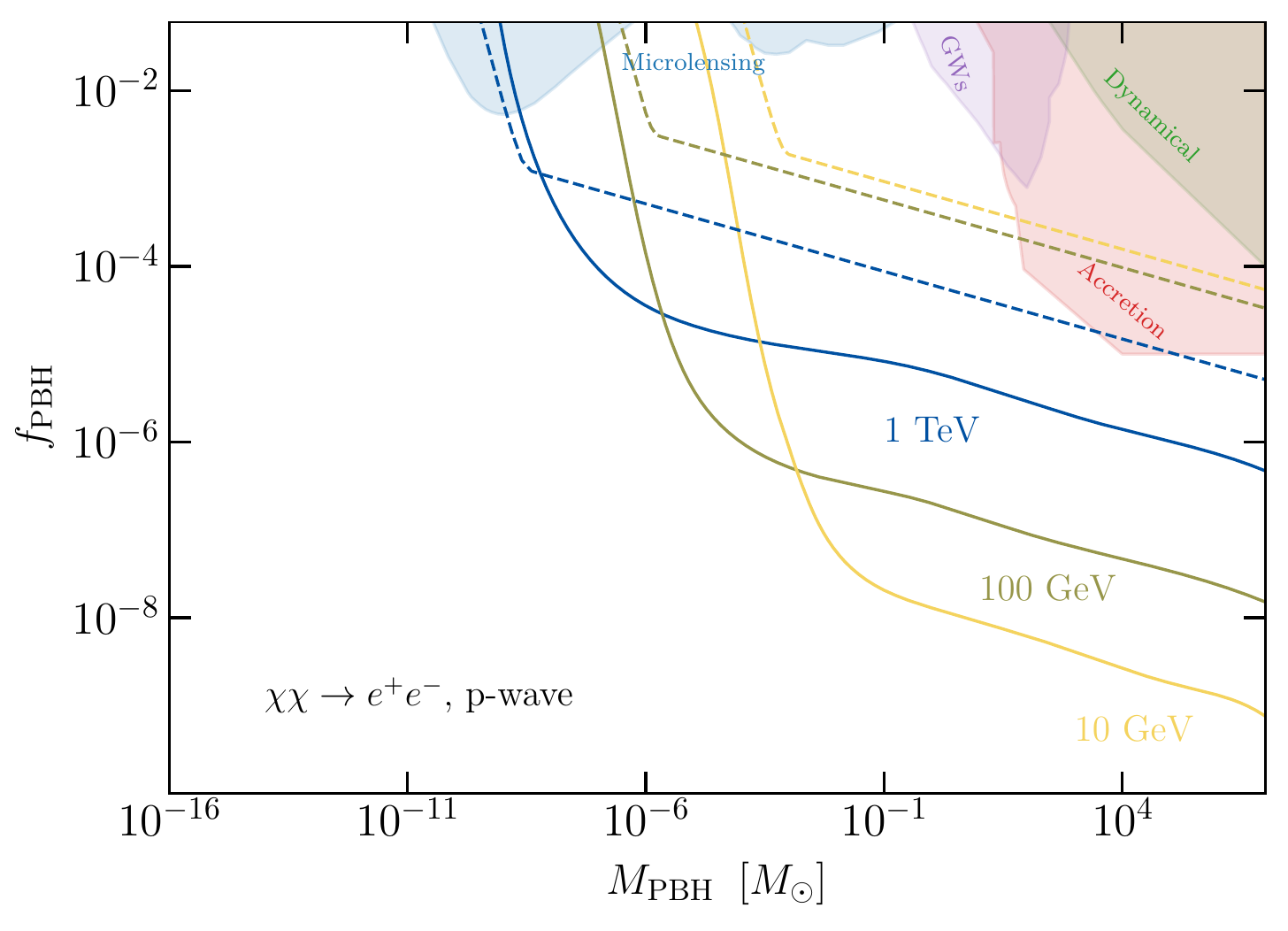}
	\caption{Top panels: The solid (dashed) lines depict the $95\%$~CL constraints on the PBH dark matter fraction as a function of the PBH mass from CMB (conservative $\gamma$-ray) observations in hybrid WIMP-PBH models for s- (left panel) and p-wave (right panel) annihilating WIMPs into the $b\bar{b}$ channel.  The cross section is assumed to be $\left\langle \sigma_A v \right\rangle=3\,\cdot\,10^{-26}/f_\chi$ $\text{cm}^3$/s and  $\left\langle \sigma_A v \right\rangle_{fo}= 3\,\cdot\,10^{-26}/f_\chi$ $\text{cm}^3$/s for the s-wave and p-wave annihilating channels, respectively.
	We illustrate three possible WIMP dark matter masses: 10 GeV, 100 GeV, and 1 TeV. The top left panel also shows the $95\%$~CL limits from the \emph{optimistic} IGRB model (dotted lines), see main text for details. Bottom left (right) panels: As in the top panels, but for s- (p-) wave annihilating WIMPs into the $e^-e^+$ channel.}
	
	\label{fig:results}
\end{figure*}

\begin{figure}
	\centering
		\includegraphics[width=1\linewidth]{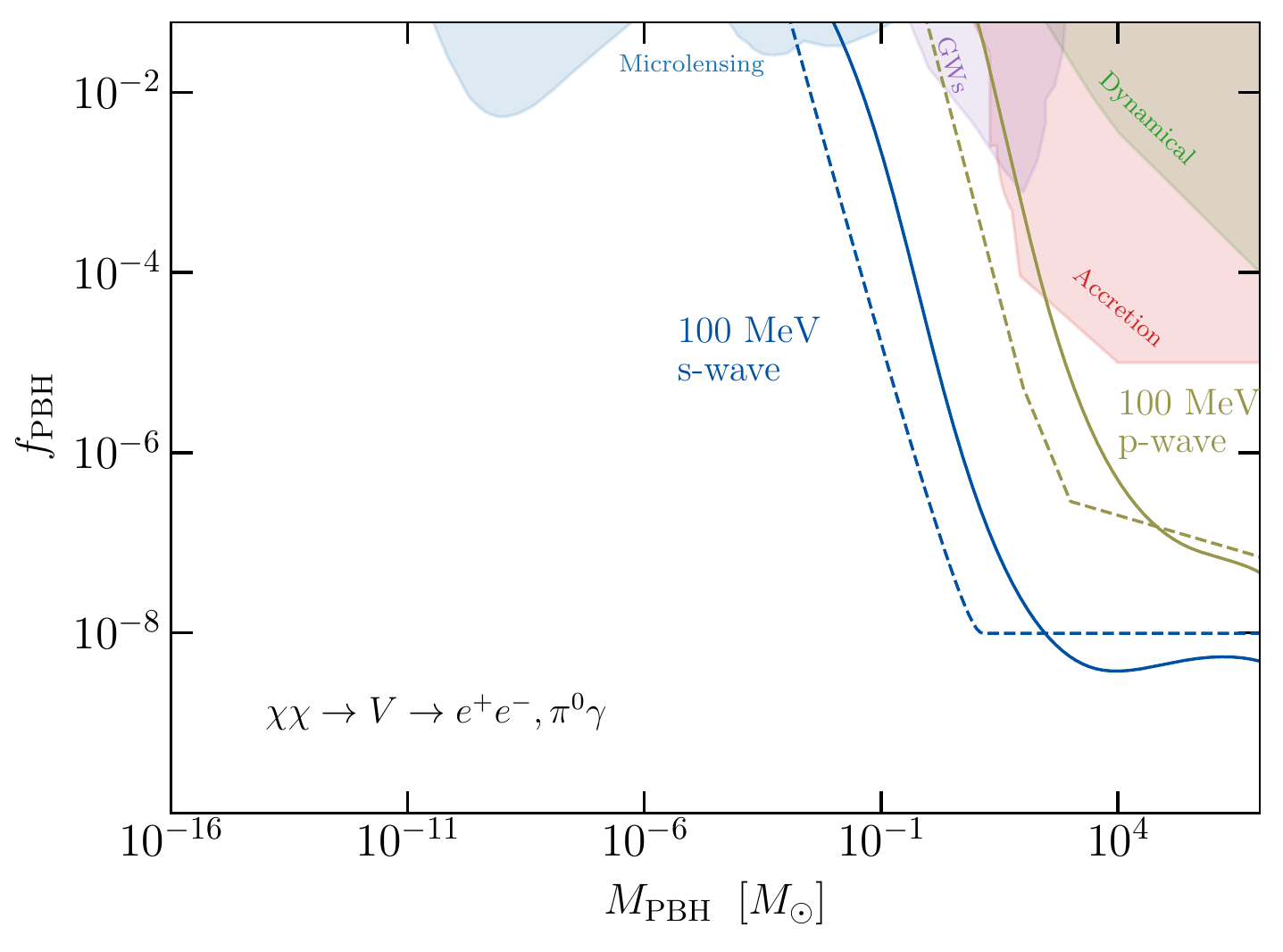}	
	\caption{Same as Fig.~\ref{fig:results}, but for the case of a 100 MeV particle annihilating through a vector portal coupling. As detailed in Sec.~\ref{sec:dm}, for this particle mass the dominant annihilation channel is $e^+ e^-$. The cross section at freeze out is assumed to be $\left\langle \sigma_A v \right\rangle= 10^{-28}/f_\chi$ $\text{cm}^3$/s for both the s-wave and p-wave cases. }
	\label{fig:results_ldm}
\end{figure}

Figures \ref{fig:results_broad}, \ref{fig:results} and \ref{fig:results_ldm} illustrate the main findings of this work, showing constraints on $f_{\rm PBH}$ as a function of $M_{\rm PBH}$ for various scenarios. Figure~\ref{fig:results_broad} illustrates the limits derived on both the monochromatic and broad log-normal (with $\sigma = 2$) PBH mass functions  for a WIMP (Majorana) dark matter particle of $10$~GeV annihilating via s-wave purely into the $b\bar{b}$ channel. At high PBH masses, the constraints are insensitive to the details of the mass function. However, at lower masses, an extended mass distribution results into constraints which are in general orders of magnitude stronger than those of the monochromatic distribution. In Fig.~\ref{fig:results_broad}, we also illustrate for comparative  purposes the constraints derived using the conservative IGRB, which are roughly two orders of magnitude weaker than those from the CMB for $M_{\rm PBH} \gtrsim 10^{-6}M_\odot$, and comparable for smaller PBH masses.  

Figure \ref{fig:results} depicts the constraints on $f_{\rm PBH}$ for
different WIMP candidates. Namely, we vary the WIMP mass, the
annihilation channel, and the velocity dependence of the annihilation
cross section. We show in the top (bottom) panel the annihilation of a
10 GeV, 100 GeV, and 1 TeV Majorana dark matter particle into the
$b\bar{b}$ ($e^+e^-$) channel, assuming a monochromatic PBH mass
function. The cases for s-wave (p-wave) annihilations are illustrated
in the left (right) panels. As before, we show the conservative IGRB
constraints using dashed lines, as well as the optimistic IGRB
constraints in the top left panel using dotted lines. We note that in
all cases, the cosmological constraints tend to be comparable to, or
stronger than, those derived using the IGRB. The strength of the
cosmological constraints is particularly pronounced for the case of
WIMPs annihilating into $e^+e^-$.  This is due to the fact that
$e^\pm$ pairs can efficiently heat and ionize the IGM, but only
generate observable $\gamma$-rays via inverse Compton scattering (ICS)
off the CMB photons. ICS generates a peak in the $\gamma$-ray spectra
at low energies, which are less constrained by $\gamma$-ray
measurements (see the figure in Appendix D). The right panels of Fig.~\ref{fig:results} illustrate that constraints on $f_{\rm PBH}$ remain relatively strong in the case of p-wave annihilating dark matter. Unlike in the case of s-wave annihilation, the p-wave constraints do not saturate to a fixed value of $f_{\rm PBH}$ at large PBH masses. 

The results of the vector portal light dark matter model for a 100 MeV dark matter particle are shown in Fig.~\ref{fig:results_ldm}. The fact that cosmological constraints are comparable to the IGRB limits implies that the \emph{optimistic IGRB} bounds would be stronger than those obtained using the CMB+BAO. For such models, one can only constrain PBH masses $M_{\scaleto{\mathrm{PBH}}{4pt}} \gtrsim 10^{-1} M_\odot$ (or $10^{2} M_\odot$ in the case of p-wave annihilating dark matter), as the kinematic suppression is increasingly strong at low dark matter masses.

It is important to point out that the constraints derived here differ notably from a number of previous works on the subject. We highlight the relative differences for either the IGRB and/or the CMB bounds in what follows.\\

{\bf Comparison with Adamek et al \cite{Adamek:2019gns}}

The optimistic IGRB limits derived in this work are weaker by a factor $\mathcal{O}(10)$ for WIMP masses in the range 100 GeV to 1 TeV, and by a factor ~$\mathcal{O}(10^2)$ for WIMP masses closer to 10 GeV with respect to those derived in Ref.~\cite{Adamek:2019gns}. 

\begin{itemize}
\item The $\mathcal{O}(10)$ difference in the 100 GeV to 1 TeV WIMP mass range is due to the use of different cosmological parameters (precise epoch and energy density at matter-radiation equality, accounting for a factor $\sim 5$) and also due to a missing factor of 2 in the denominator of the formula for the annihilation rate of self-conjugated WIMPs. Additionally, our limits also include the  factor of $\sim$1.5 which is introduced by correcting for the redshift independent annihilation rate (as mentioned above). 

\item The remaining ~$\mathcal{O}(10)$ difference at 10 GeV is due to the choice of the limit on $\Gamma_{\rm dec}$, which is taken from Fig. 3 of \cite{Ando:2015qda}. While the authors of \cite{Adamek:2019gns} adopt a single mass-independent limit of $10^{28}$ s (which is technically only valid for WIMP masses near 100 GeV), we have appropriately scaled the data to be valid at all WIMP masses. 
\end{itemize}

{\bf Comparison with Carr et al \cite{Carr:2020mqm}}
\newline
The authors of \cite{Carr:2020mqm} follow a slightly different approach. They compare the theoretical $\gamma$-ray signal today integrated for all energies above 100 MeV  with a sky-integrated flux threshold of $\Phi_{\rm res} \sim 10^{-7} \, {\rm cm^{-2} \, s^{-1}}$.

Compared to our \emph{optimistic IGRB} case, the constraints of \cite{Carr:2020mqm} are more stringent by one, two and four orders of magnitude for WIMP masses of 1000, 100 and 10 GeV respectively. These differences are mostly driven by their choice of the upper limit on the flux, $\Phi_{\rm res}$, which is approximately four orders of magnitude below the observed energy-integrated IGRB~\cite{Fermi-LAT:2014ryh} -- this difference is presumably attributed to the subtraction of the astrophysical background, however it is unclear how such a background subtraction would be achieved. In addition, our results have a smaller dependence on the WIMP mass (\eg our PBH limits for 1 TeV and 100 GeV are almost identical, while for \cite{Carr:2020mqm} they differ by about one order of magnitude). This dependence is a consequence of using a limit to the flux integrated in the entire $\textit{Fermi}$-LAT energy band instead of using a binned analysis. 

There are other factors which also contribute to the difference, albeit to a lesser extent. These include a missing factor of $\frac{1}{2}$ in the annihilation rate, and the fact that we use the latest \textit{Planck} 2018 cosmological values for the parameters, as, for instance, for $t_{\mathrm{eq}}$ and $\rho_{\mathrm{eq}}$.

{\bf Comparison  with Tashiro et al \cite{Tashiro:2021xnj}}
\newline
The most recent CMB constraints on the hybrid PBH-WIMP scenario have been derived in \cite{Tashiro:2021xnj}. This study, however, focused exclusively on the high PBH mass region, where the kinetic energy of the WIMPs can be neglected. The results of Ref.~\cite{Tashiro:2021xnj} on the $e^\pm$ channel do not differ significantly from the results presented here, however we find discrepancies in the $b\bar{b}$ channel which lead to disagreements of up to one order of magnitude. There are a number of potential reasons for this discrepancy, as, for instance, differences in the different methodology employed. Namely,  Ref.~\cite{Tashiro:2021xnj} uses the so-called ``on-the-spot'' approximation, which assumes that all energy is deposited at the redshift of injection. Furthermore, this approximation relies on simplified redshift averaged fitting functions to estimate how energy is deposited into the system.

\section{Conclusions}\label{sec:conc}

The discovery of gravitational waves from the coalescence of binary systems of black holes has revived the interest in scenarios where the Cold Dark Matter (CDM) can be in the form of Primordial Black Holes (PBHs). While PBHs with masses $M_{\scaleto{\mathrm{PBH}}{4pt}} \gtrsim 10^{-11} \, M_\odot$ cannot comprise the entirety of dark matter, the presence of even a small abundance of such objects can have profound consequences. These objects can efficiently accrete large amounts of the primary dark matter component in the early Universe, forming high density spikes. If the major dark matter component is due to WIMPs, the high densities achieved in these spikes dramatically enhance the WIMP annihilation rate, leading therefore to strong observable effects in astrophysics and cosmology. 

 Here, we revisit cosmological and astrophysical constraints on the hybrid WIMP-PBH dark matter scenario. This work offers a major improvement with respect to previous cosmological analyses, including the latest observational data, a proper treatment of energy deposition and propagation in the IGM, a WIMP density profile incorporating the kinematic suppression arising at low WIMP/PBH masses, and a through investigation into a broad array of scenarios (including extended PBH mass functions, s- and p-wave annihilations, and a wide variety of WIMP models).  For comparison, we re-derive the astrophysical constraints from gamma-ray observations of the IGRB --  importantly, we find that previous results have  notably overestimated the constraints on the PBH fraction from these measurements (in some cases by many orders of magnitude).

For most scenarios we find that the cosmological constraints from CMB and BAO observations are comparable to, or even slightly stronger than those derived from the IGRB. Despite the fact that our results differ qualitatively from those previously derived in the literature, the conclusions are roughly the same: if WIMPs comprise the primary component of dark matter, PBHs are severely constrained from contributing notably to the dark matter density. Exceptions remain, however, for PBH masses $M_{\scaleto{\mathrm{PBH}}{4pt}} \lesssim 10^{-6} \, M_\odot$, and very light p-wave annihilating dark matter. 
\begin{acknowledgments}
OM is supported by the Spanish grants PID2020-113644GB-I00, PROMETEO/2019/083 and by the European ITN project HIDDeN (H2020-MSCA-ITN-2019//860881-HIDDeN). SJW is supported by the European Research Council (ERC) under the European Union's Horizon 2020 research and innovation programme (Grant agreement No. 864035 - Un-Dark). This research was supported by the Excellence Cluster ORIGINS which is funded by the Deutsche Forschungsgemeinschaft (DFG, German Research Foundation) under Germany's Excellence Strategy EX--2094 390783311.

\end{acknowledgments}

\setcounter{equation}{0}
\setcounter{figure}{0}
\setcounter{table}{0}
\setcounter{section}{0}
\makeatletter
\renewcommand{\theequation}{S\arabic{equation}}
\renewcommand{\thefigure}{S\arabic{figure}}
\renewcommand{\thetable}{S\arabic{table}}

\onecolumngrid
\clearpage
\appendix

\begin{center}
  \textbf{\large Supplementary Material}\\[.2cm]
  \vspace{0.05in}
  
\end{center}

We present below the main formulas we have used to compute the annihilation rates for the cases of s-wave and p-wave annihilating WIMPs with a monochromatic PBH mass function, and s-wave annihilation using a log-normal mass function.  We also provide illustrative examples of the annihilation spectra with comparison to the observed the extra-galactic $\gamma$-ray background.

\section{S-wave annihilation rate for monochromatic PBH spectrum}
\label{sec:apb}

The annihilation rate for s-wave annihilating dark matter is dominated by the largest radii for which $\rho(r) = \rho_{\rm max}$, which we denote $r_{\rm cut}$ (\ie the radius at which the profile is cut, or truncated, by the presence of annihilations). We can directly solve for the value of $r_{\rm cut}$ by equating $\rho_{\rm max}$ to the piece-wise function given in Eq.~\ref{rho_chi}. The value of $r_{\rm cut}$ in this case is given by\\
\begin{equation} \label{eq:rcut}
\begin{split}
r_{\rm cut} = \left\{
\begin{array}{ll}
 \left(f_\chi \frac{\rho_{\mathrm{eq}}}{2} \right)^{2/3}\left(\frac{M_{\mathrm{PBH}}}{M_\odot}\right)\hat{r}\rho_{\mathrm{max}}^{-2/3} & \mathrm{for\ }   f_\chi \rho_{\scaleto{\mathrm{KD}}{4pt}} > \rho_{\rm max} \ge \rho_{\scaleto{\mathrm{K}}{4pt}}  ,\\
 \left(f_\chi \frac{\rho_{\mathrm{eq}}}{2} \right)^{4/9}\left(\frac{M_{\mathrm{PBH}}}{M_\odot}\right)^{1/3}\bar{r}\rho_{\mathrm{max}}^{-4/9} & \mathrm{for\ } \rho_{\rm max} < \rho_{\scaleto{\mathrm{K}}{4pt}},\\
\end{array}
\right.
\end{split}
\end{equation}\\
\noindent with $\rho_{\scaleto{\mathrm{K}}{4pt}}$ given by
\begin{equation} \label{eq:rcut}
\rho_{\scaleto{\mathrm{K}}{4pt}} = f_\chi \frac{\rho_{\mathrm{eq}}}{2} (G M_{\mathrm{PBH}})^3 \left(\frac{t_\mathrm{eq}}{2}\right)^{3/2} \left(\frac{m_\chi}{t_{\scaleto{\mathrm{KD}}{4pt}}T_{\scaleto{\mathrm{KD}}{4pt}}}\right)^{9/2} \, .
\end{equation}
Note that in defining Eq.~\ref{eq:rcut} we have neglected the possibility that $\rho_{\rm max} > f_\chi \rho_{\rm KD}$, as this never occurs for the models of interest.

The annihilation rate can then be directly calculated by integrating the  density profile; the result is given by\\
\begin{equation} \label{ann_rate0}
\begin{split}
\Gamma_{\mathrm{ann}} = \left\{
\begin{array}{ll}
\frac{2^{2/3} \pi \left\langle \sigma_A v \right\rangle^{1/3}G M_{\mathrm{PBH}} t^2_{\mathrm{eq}} (f_x \rho_{\mathrm{eq}})^{4/3} }{ m_\chi^{4/3} t^{2/3}} + \Upsilon~;
&  r_{\mathrm{cut}} \ge r_K, r_{\mathrm{cut}} > r_S \\
\frac{\left\langle \sigma_A v \right\rangle m_\chi \pi (f_\chi \rho_{\mathrm{eq}})^2 (G M_{\mathrm{PBH}}t_{\mathrm{eq}})^3 }{2 (t_{\mathrm{KD}} T_{\mathrm{KD}})^3}  \left(1 +  \mathrm{Log}\left[\frac{2^{ 5/3} (t_{\mathrm{KD}} T_{\mathrm{KD}})^3 }{  (f_\chi \rho_{\mathrm{eq}} \left\langle \sigma_A v \right\rangle t)^{2/3 } (GM_{\mathrm{PBH}})^2 t_{\mathrm{eq}} m_\chi^{7/3} }\right] \right) + \Upsilon ~,
& r_{\mathrm{cut}} < r_K, r_{\mathrm{cut}} > r_S\\
\end{array}
\right.
\end{split}
\end{equation}
where 
\begin{equation} \label{ann_rate2}
\Upsilon= -\frac{16 \pi(G M_{\mathrm{PBH}})^3}{3 \left\langle \sigma_A v \right\rangle t^2} \, .
\end{equation}
The factor $\Upsilon$ accounts for the fact that there is no injected energy from annihilations within the Schwartschild radius, and is typically sufficiently small to be neglected.

\section{P-wave annihilation rate for monochromatic PBH spectrum}
\label{sec:apc}

\begin{figure*}
	\centering
        \includegraphics[width=0.49\linewidth]{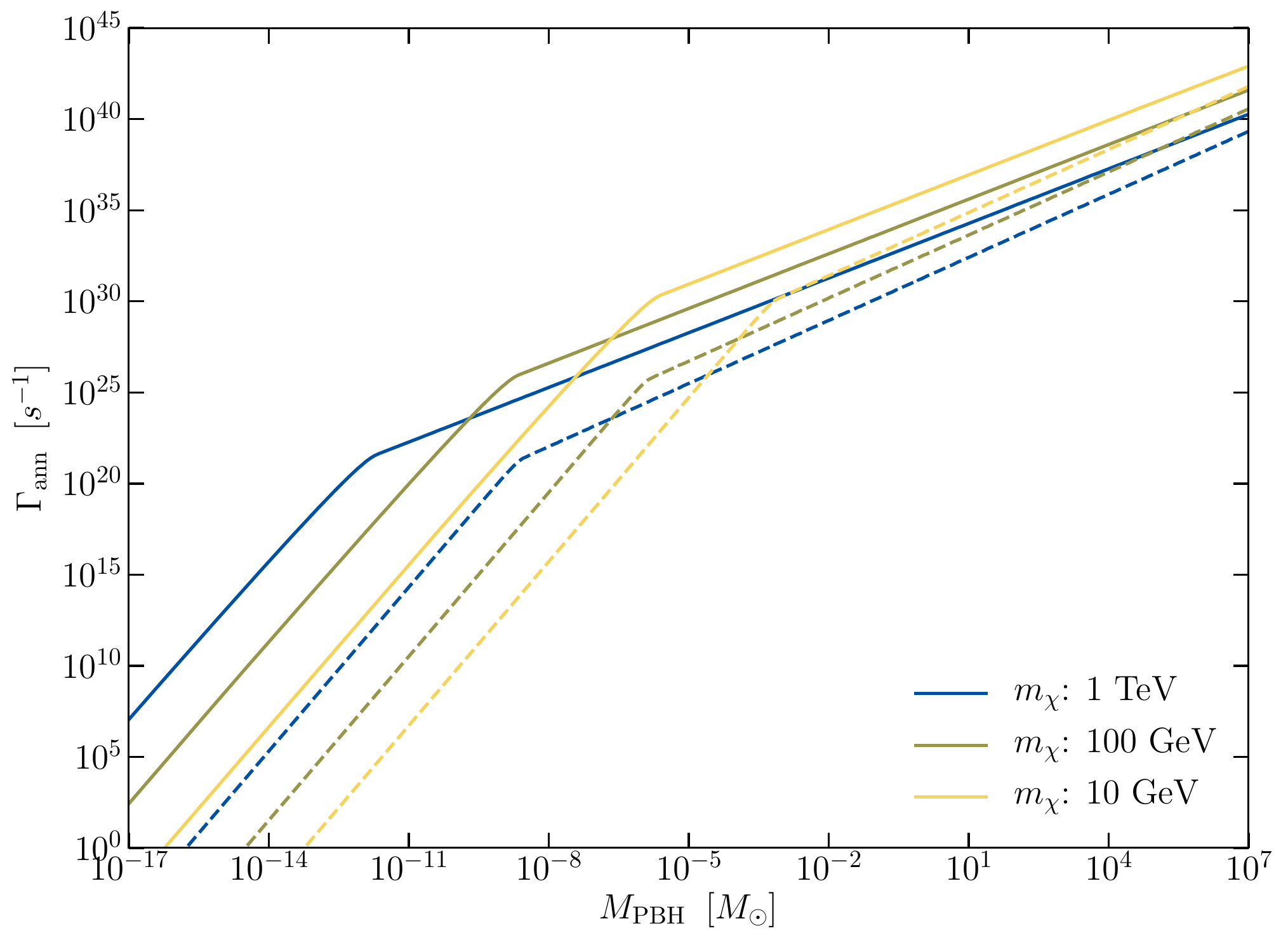}
	    \includegraphics[width=0.49\linewidth]{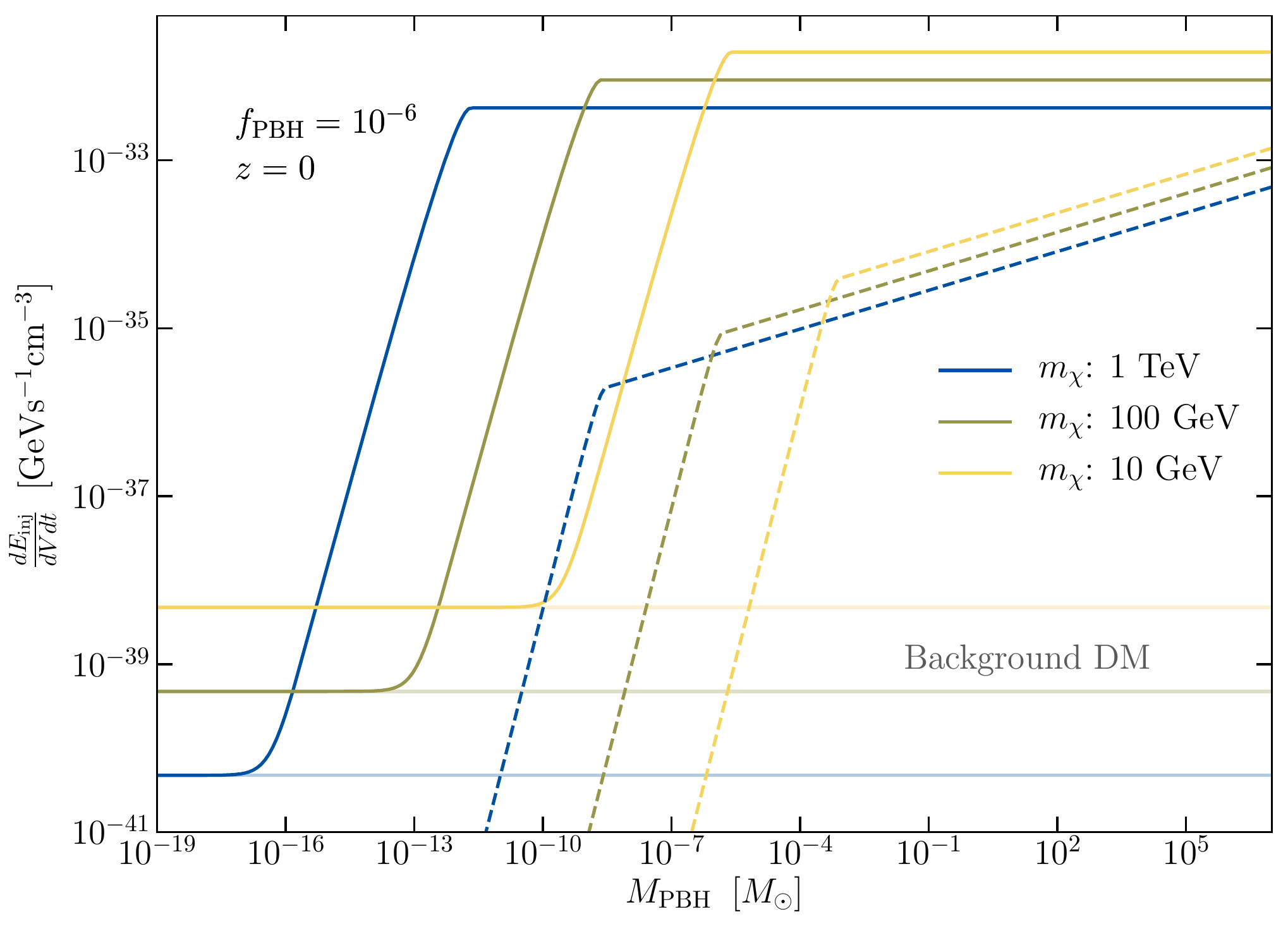}	
	\caption{The left (right) panels depict the annihilation rate per mini-halo and the diffuse injected power density for a PBH fraction $10^{-6}$ at present ($z=0$) respectively as a function of the PBH mass for three different DM masses and also for the cases of s-wave (solid lines) and p-wave (dashed lines) annihilations.}
	\label{fig:rho2}
\end{figure*}

The case of p-wave annihilation differs slightly from the case above owing to the fact that $\rho_{\rm max}$ varies radially. Nevertheless, the dominant contribution from WIMP annihilations still comes from the largest radii for which $\rho_{\rm max}(r) = \rho(r)$, and thus can be solved in a comparable manner. In this case, one finds  $r_{\rm cut}$ is given by
\begin{equation} \label{eq:rcut}
\begin{split}
r_{\rm cut} = \left\{
\begin{array}{ll}
  (f_\chi \frac{\rho_{\mathrm{eq}}}{2})^{ 2/5 } \left( \frac{t_{\mathrm{eq}} }{t_{\scaleto{\mathrm{KD}}{4pt}} T_{\scaleto{\mathrm{KD}}{4pt}}}  \right)^{3/5} \left(\frac{t \left\langle \sigma_A v \right\rangle_{\mathrm{fo}} }{v_{\mathrm{fo}}^2}\right)^{2/5} (G M_{\mathrm{PBH}}) m_\chi^{1/5} & \mathrm{for\ }   f_\chi \rho_{\scaleto{\mathrm{KD}}{4pt}} < \rho_{\rm max}(r_{\rm cut}) \leq \rho_{\scaleto{\mathrm{K}}{4pt}}  ,\\
  \left[f_\chi \frac{\rho_{\mathrm{eq}}}{2^{ 1/4 }}(G M_{\mathrm{PBH}})^{7/4}  t_{\mathrm{eq}}^{3/2}\frac{\left\langle \sigma_A v \right\rangle_{\mathrm{fo}} t}{m_\chi v_{\mathrm{fo}}^2 }\right]^{4/13} & \mathrm{for\ } \rho_{\rm max}(r_{\rm cut}) > \rho_{\scaleto{\mathrm{K}}{4pt}},\\
\end{array}
\right.
\end{split}
\end{equation}\\
\noindent with $\rho_{\scaleto{\mathrm{K}}{4pt}}$ as defined in Appendix~\ref{sec:apb}. The p-wave annihilation rate is then defined as follows
\begin{equation} \label{ann_rate0}
\begin{split}
\Gamma_{\mathrm{ann}} = \frac{1}{2 m_\chi^2} \int_V dV \left\langle \sigma_A v \right\rangle^{\mathrm{p-wave}} \rho_x^2 = \left\{
\begin{array}{ll}
\Gamma_{\mathrm{ann}}'~; &  r_{\mathrm{cut}} \ge r_K, r_{\mathrm{cut}} > r_S \\
\Gamma_{ann}''~, & r_{\mathrm{cut}} < r_K, r_{\mathrm{cut}} > r_S
\end{array}
\right.
\end{split}
\end{equation}\\

where the annihilation rates $\Gamma_{\mathrm{ann}}'$ and $\Gamma_{\mathrm{ann}}'$ are given by

\begin{equation} \label{ann_rate2_pw}
\begin{split}
\Gamma_{\mathrm{ann}}'= \frac{\pi v_{\mathrm{fo}}^2}{20\left\langle \sigma_A v \right\rangle_{\mathrm{fo}}G M_{\mathrm{PBH}} t^2 } \left((5+2^{13/6})\left(\frac{f_\chi (G M_{\mathrm{PBH}})^{7/4} \rho_{\mathrm{eq}}\left\langle \sigma_A v \right\rangle_{\mathrm{fo}} t \cdot t_{\mathrm{eq}}^{3/2}}{m_\chi v_{\mathrm{fo}}^2} \right)^{16/13}-160 (G M_{\mathrm{PBH}})^4\right)~;
\end{split}
\end{equation}

\noindent and

\begin{equation} \label{ann_rate4_pw}
\begin{split}
\Gamma_{\mathrm{ann}}''=\frac{(GM_{\mathrm{PBH}})^3\pi}{40 \left\langle \sigma_A v \right\rangle_{\mathrm{fo}} v_{\mathrm{fo}}^2 t^2 (t_{\mathrm{KD}}T_{\mathrm{KD}})^6}\left(-6(f_\chi G M_{\mathrm{PBH}} \rho_{\mathrm{eq}} \left\langle \sigma_A v \right\rangle_{\mathrm{fo}} t)^2 (m_\chi t_{\mathrm{eq}})^3 t_{\mathrm{KD}}T_{\mathrm{KD}} \right.\\
- \left. 320(t_{\mathrm{KD}}T_{\mathrm{KD}})^6v_{\mathrm{fo}}^4 + 25\cdot 2^{2/5}m_\chi^{4/5}(f_\chi \rho_{\mathrm{eq}})^{8/5} t_{\mathrm{eq}} (t_{\mathrm{KD}} T_{\mathrm{KD}})^{18/5} t_{\mathrm{eq}}^{7/5} (\left\langle \sigma_A v \right\rangle_{\mathrm{fo}} t)^{8/5} v_{\mathrm{fo}}^{4/5}\right)~.
\end{split}
\end{equation}
Fig.\ref{fig:rho2} compares the annihilation rate (left) and the
injected power per unit volume (right) for s-wave and the p-wave
annihilations. As expected, p-wave annihilations are significantly
suppressed with respect to s-wave (although the p-wave annihilation
rate tends toward s-wave in the limit of heavier PBHs), and the
annihilation rate is only enhanced with respect to the contribution
from the background component for sufficiently large PBH masses.

\section{S-wave annihilation rate for broad PBH mass spectrum}
\label{sec:apd}

In this work we explore the sensitivity of cosmological and astrophysical bounds to the choice of PBH mass function. In particular, we compare the constraints using the monochromatic mass function to that of a log-normal mass function \cite{Carr:2017jsz} (see Eq.~\ref{psi}  of Sec.~\ref{sec:mass}). The annihilation rate of the log-normal spectrum is given by
\begin{equation} \label{ann_rate_broad}
\begin{split}
\Gamma_{\mathrm{ann,b}} \equiv
\int dM \psi(M) \int_V \frac{\left\langle \sigma_A v \right\rangle}{2 m_\chi^2} \rho_\chi^2(M)=
\Gamma_{\mathrm{ann,b}}^1 + \Gamma_{\mathrm{ann,b}}^2~;
\end{split}
\end{equation}
 with
\begin{equation} \label{ann_rate_broad3}
\begin{split}
\Gamma_{\mathrm{ann,b}}^1 = \int_{M_{\mathrm{min}}}^{M_{\mathrm{cut}}(t)} dM \psi(M) \Gamma_{\mathrm{ann}}''(M) = \Upsilon_1 + \Upsilon_2 + \Upsilon_4~;\\
\Gamma_{\mathrm{ann,b}}^2 = \int_{M_{\mathrm{cut}}(t)}^{M_{\mathrm{max}}} dM_i \psi(M_i) \Gamma_{\mathrm{ann}}'(M) = \Upsilon_3 + \Upsilon_5 \, .
\end{split}
\end{equation}
Here, we have defined $M_{\mathrm{cut}}(M, t)$ via $r_K(M_{\mathrm{cut}}) = \bar{r}_{\mathrm{cut}}(M_{\mathrm{cut}})$, which yields a solution 
\begin{equation} \label{M_cut}
\begin{split}
M_{\mathrm{cut}}(M, t) = 2^{5/6} \left(\frac{T_{\mathrm{KD}}t_{\mathrm{KD}}}{m_\chi}\right)^{3/2} \left(f_\chi \rho_{eq}\right)^{-1/3}G^{-1}t^{-1/2}_{eq} \left(\frac{m_\chi}{\left\langle \sigma_A v \right\rangle t}\right)^{1/3},\\
\end{split}
\end{equation}
with $\Upsilon_i$ given by
\begin{eqnarray} \label{ann_rate_broad2_2}
\Upsilon_1&=&-\frac{\mathrm{Exp}[\frac{9}{2}\sigma^2] f_x^2 G^3 M_{\mathrm{pk}}^3 m_\chi \pi \rho_{\mathrm{eq}}^2 \left\langle \sigma_A v \right\rangle t_{\mathrm{eq}}^3}{4(t_{\mathrm{KD}} \cdot T_{\mathrm{KD}})^3}\left(\mathrm{Erf}[\Pi_2] - \mathrm{Erf}[\Pi_3]\right)(1 + 2\mathrm{Log}[\Pi_1 M_{\mathrm{pk}}])~;\\ \nonumber
\Upsilon_2&=&-\frac{f_x^2 G^3 m_\chi \pi \rho_{\mathrm{eq}}^2 \left\langle \sigma_A v \right\rangle t_{\mathrm{eq}}^3}{(t_{\mathrm{KD}} \cdot T_{\mathrm{KD}})^3}\left( \frac{1}{\sqrt{2\pi}}\mathrm{Exp}\left[-\frac{\mathrm{Log}[\frac{M_{\mathrm{min}}}{M_{\mathrm{pk}}}]^2}{2\sigma^2}\right]M_{\mathrm{min}}^3\sigma \right. \\ \nonumber
&-& \left.\frac{4 \sigma (t_{\mathrm{KD}} \cdot T_{\mathrm{KD}})^{9/2}}{m_\chi^{7/2}f_\chi G^3\sqrt{\pi}\rho_{\mathrm{eq}} \left\langle \sigma_A v \right\rangle t_{\mathrm{eq}}^{3/2}t}\mathrm{Exp}\left[-\frac{\mathrm{Log}[\Pi_1]^2}{2\sigma^2}\right] \right.\\
\Bigg. 
&-&\frac{1}{2}\mathrm{Exp}\left[\frac{9}{2}\sigma^2\right] M_{\mathrm{pk}}^3 (3\sigma^2+\mathrm{Log}[M_{\mathrm{pk}}]) (\mathrm{Erf}[\Pi_2] - \mathrm{Erf}[\Pi_3]) \Bigg)   \\
\Upsilon_3&=&\frac{\mathrm{Exp}\left[\frac{9}{2}\sigma^2\right]G M_{\mathrm{pk}}\pi(f_x \rho_{\mathrm{eq}})^{4/3} \left\langle \sigma_A v \right\rangle^{1/3} t_{\mathrm{eq}}^2}{2^{1/3}t^{2/3} m_\chi^{4/3}}\left(\mathrm{Erf}\left[\frac{- \sigma^2+\mathrm{Log}[\frac{M_{\mathrm{max}}}{M_{\mathrm{pk}}}]}{\sqrt{2}\sigma}\right] - \mathrm{Erf}\left[\frac{- \sigma^2+\mathrm{Log}[\Pi_1]}{\sqrt{2}\sigma}\right]\right)~;\\
\Upsilon_4&=&\frac{8\mathrm{Exp}[\frac{9}{2}\sigma^2](G M_{\mathrm{pk}})^3\pi}{3 \left\langle \sigma_A v \right\rangle t^2}\left(\mathrm{Erf}[\Pi_2] - \mathrm{Erf}[\Pi_3]\right)~;\\
\Upsilon_5&=&\frac{8\mathrm{Exp}[\frac{9}{2}\sigma^2](G M_{\mathrm{pk}})^3\pi}{3 \left\langle \sigma_A v \right\rangle t^2}\left(\mathrm{Erf}[\Pi_3] - \mathrm{Erf}\left[\frac{-3 \sigma^2+\mathrm{Log}[\frac{M_{\mathrm{max}}}{M_{\mathrm{pk}}}]}{\sqrt{2}\sigma}\right]\right) \, .
\end{eqnarray}
For simplicity we have introduced the following functions:
\begin{eqnarray} \label{ann_rate_broad2_2}
\Pi_1 & \equiv &\frac{2^{5/6}
(t_{\mathrm{KD}}\cdot T_{\mathrm{KD}})^{3/2} }{m_\chi^{7/6}G M_{\mathrm{pk}} (f_\chi
\rho_{\mathrm{eq}}\left\langle \sigma_A v \right\rangle t)^{1/3} \sqrt{t_{\mathrm{eq}}}} \, \\ \Pi_2 & \equiv &\frac{-3 \sigma^2+\mathrm{Log}[\frac{M_{\mathrm{min}}}{M_{\mathrm{pk}}}]}{\sqrt{2}\sigma} \, \\ \Pi_3 &\equiv &\frac{-3 \sigma^2+\mathrm{Log}[\Pi_1]}{\sqrt{2}\sigma} \, .
\end{eqnarray}
 The factors $\Upsilon_4$ and $\Upsilon_5$ 
account for the fact there is no injected energy from annihilations within the Schwartschild radius, and are typically $<\mathcal{O}(10^{-4})$ of the total annihilation rate for most of the parameter space explored in this work. The parameters  $M_{\mathrm{max}}$ and $M_{\mathrm{min}}$ are the maximum and minimum PBH masses of the model ($10^{10}$ and $10^{-30}$ solar masses in our calculations). Note that the mean UCMH mass, $\overline{M}_{\scaleto{\mathrm{UCMH}}{4pt}}$ scales as $\overline{M}_{\scaleto{\mathrm{UCMH}}{4pt}}(z)=\overline{M}_{\scaleto{\mathrm{PBH}}{4pt}}(1+z_{\mathrm{eq}})({1+z})^{-1}$ due to secondary accretion, with $\overline{M}_{\scaleto{\mathrm{PBH}}{4pt}}(t)$ is defined as
\begin{equation} \label{mean_PBH_mass}
\begin{split}
\overline{M}_{\scaleto{\mathrm{PBH}}{4pt}}(t) = 
\int_{0}^{\infty} M \psi(M)dM~.
\end{split}
\end{equation}\\

\section{Extra-galactic $\gamma$-rays from dark matter annihilation}
\label{sec:ape}

In what follows, we adopt the method of Ref.~\cite{Cirelli:2010xx} to compute the flux of extra-galactic $\gamma$-rays from dark matter annihilations  at $z=0$. The gamma ray flux reads as:

\begin{equation} \label{egrb}
\frac{d \Phi_{\mathrm{EG}_\gamma}}{dE_\gamma}(E_\gamma,)=\frac{1}{E_\gamma}\int_0^\infty dz^\prime\frac{1}{H(z^\prime)(1+z^\prime)^4} j_{\text{EG}_\gamma}(E^\prime_\gamma, z^\prime)e^{-\tau(E_\gamma, z^\prime)}~,
\end{equation}
with $H(z)$ the Hubble expansion rate and the emissivity $j_{\mathrm{EG}_\gamma}$
\begin{equation} \label{egrb2}
j_{\mathrm{EG}_\gamma}(E^\prime_\gamma, z^\prime)=j_{\mathrm{EG}_\gamma}^{\mathrm{prompt}}(E^\prime_\gamma, z^\prime)+j_{\mathrm{EG}_\gamma}^{\mathrm{IC}}(E^\prime_\gamma, z^\prime)~,
\end{equation}
 accounts for the contribution from both a prompt $\gamma$-ray signal and from $\gamma$-rays from Inverse Compton Scattering (ICS):
\begin{eqnarray} \label{egrb3}
j_{\mathrm{EG}_\gamma}^{\mathrm{prompt}}(E^\prime_\gamma, z^\prime)&=&E^\prime_\gamma\
\frac{1}{2}B(z')\left(\frac{\bar{\rho}(z')}{m_\chi}\right)^2\sum_f \left\langle \sigma_A v \right\rangle_f \frac{dN_\gamma^f}{dE_\gamma}(E_\gamma^\prime)~\\\nonumber
j_{\mathrm{EG}_\gamma}^{IC}(E^\prime_\gamma, z^\prime)&=&
2\int_{m_e}^{m_\chi}dE_e\frac{\mathcal{P}_{\mathrm{IC}}^{\mathrm{CMB}}(E^\prime_\gamma, E_e, 0)}{b_{\mathrm{IC}}^{\mathrm{CMB}}(E_e, 0)}\int_{E_e}^{m_\chi} d\tilde{E}_e\frac{dN_e}{d\tilde{E}_e}(E_\gamma^\prime)\frac{1}{2}B(z')\left(\frac{\bar{\rho}(0)}{m_\chi}\right)^2(1+z')^3\sum_f \left\langle \sigma_A v \right\rangle_f \, .
\end{eqnarray}
 In the above equations, $\bar{\rho}$ represents the average cosmological dark matter density, $dN_\gamma/dE_\gamma$ the spectrum of prompt photons, $B$ the boost factor and we have defined
\begin{eqnarray} \label{egrb4}
\mathcal{P}_{\text{IC}}^{\mathrm{CMB}}(E^\prime_\gamma, E_e, 0)&=&\frac{3\sigma_T}{4 \gamma^2}E_\gamma\int_0^1 dy\frac{n_{\mathrm{CMB}}(E_\gamma^0(y), 0)}{y}[2 y \text{Log} (y)+y+1-2y^2]~\\\nonumber
b_{\text{IC}}^{\mathrm{CMB}}(E_e, 0)&=&\frac{4\sigma_T}{3m_e^2}E_e^2 u_{\gamma,\mathrm{CMB}}(0)R_{\mathrm{CMB}}^{\mathrm{KN}}(E_e) \, .
\end{eqnarray}
Here,  $y=E_\gamma/(4 \gamma^2E_\gamma^0)$, $u_{\gamma,\rm CMB}(0)=0.260\; \text{eV}/\text{cm}^3$, $R_{\mathrm{CMB}}^{\mathrm{KN}}(E_e)=1$ and $n_\text{CMB}$ is the number density of CMB photons per unit energy ($\text{cm}^{-3} \text{eV}^{-1}$).

\begin{figure*}
	\centering
		\includegraphics[width=.49\linewidth]{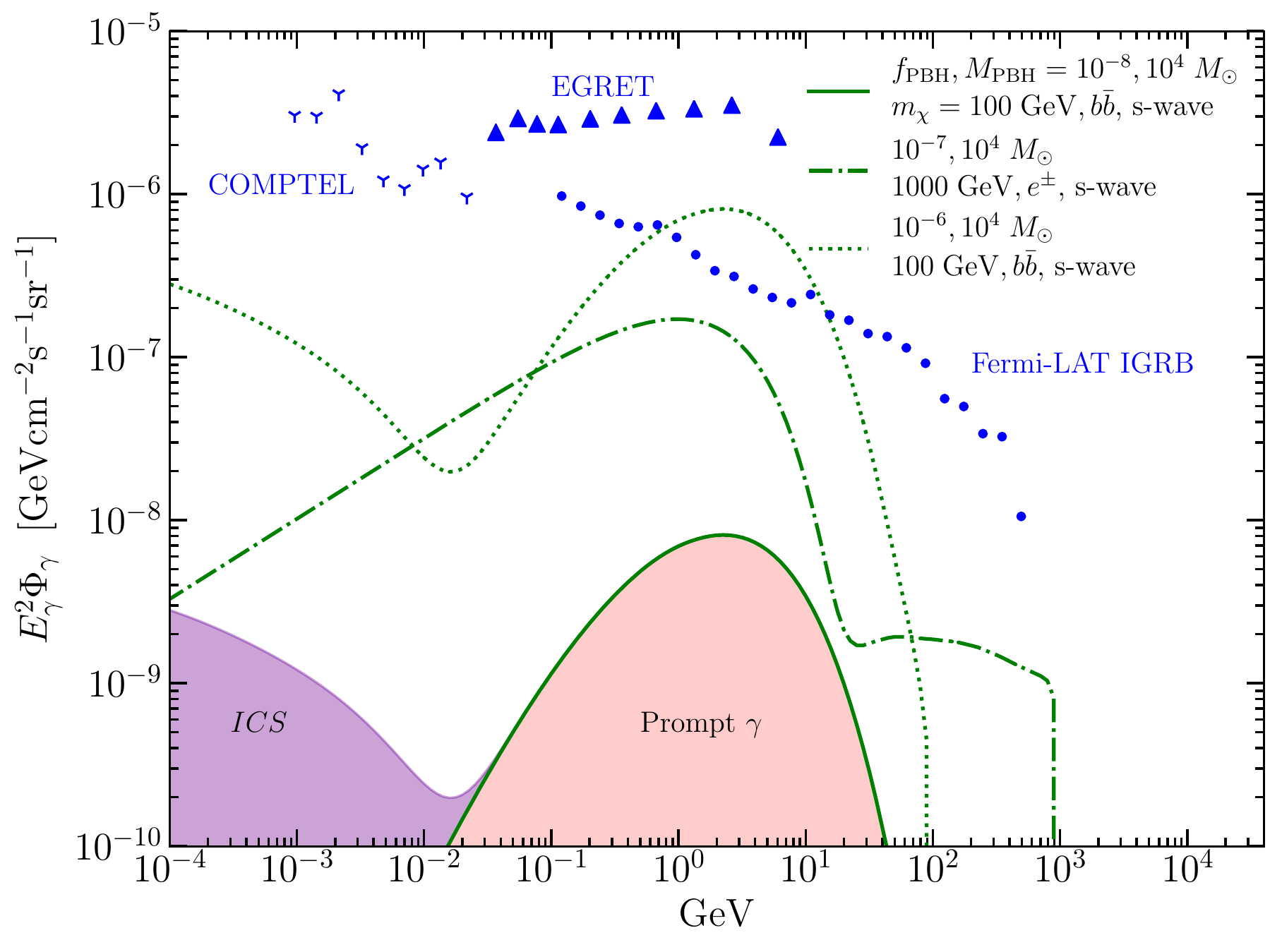}	
		\includegraphics[width=.49\linewidth]{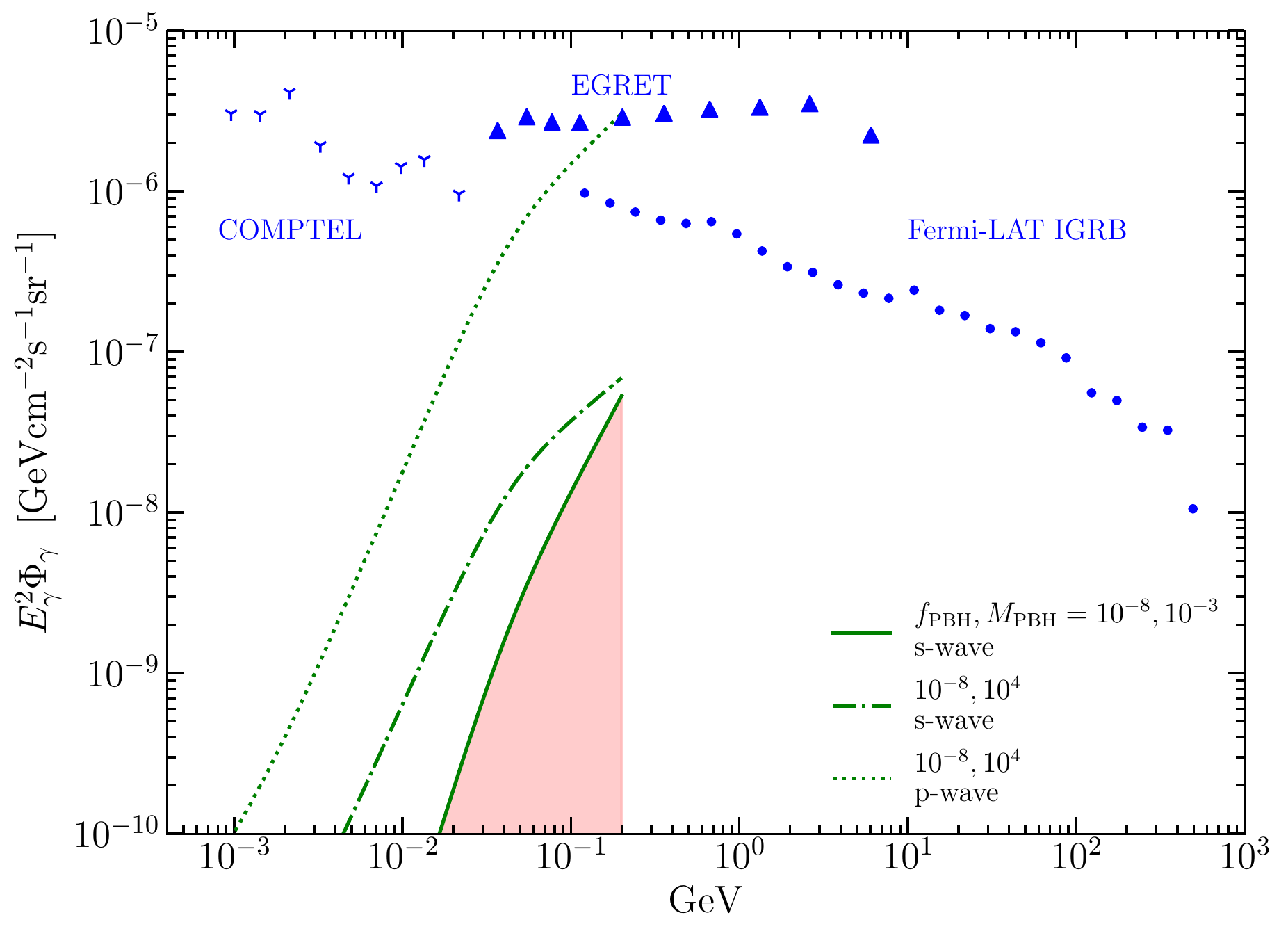}
	\caption{Left panel: Gamma ray spectrum from DM annihilations into the $e^{+}e^{-}$ and $b\bar{b}$ channels for several PBH and dark matter particle masses. For the case $m_\chi=100$~GeV and the $b\bar{b}$ channel, we illustrate separately the contribution from the ICS and prompt gamma ray fluxes (purple and green shaded regions respectively). The data points depict the observations from COMPTEL, EGRET, Fermi-LAT. Right panel: Gamma ray spectrum from DM annihilations within the light dark matter model ($m_\chi= 100$~MeV) considered here into the $e^{+}e^{-}$ channel. Several PBH masses and fractions are shown for both s-wave and p-wave annihilating channels. The contribution from gamma rays from ICS is negligible in this model, see text for details.}
	\label{fig:gamma_refs}
\end{figure*}

Figure~\ref{fig:gamma_refs} shows the $\gamma$-ray signal for the two different WIMP annihilation  models.  Notice from the right panel, corresponding to the low dark matter masses, that, in general, the energy from annihilations into $e^\pm$ pairs is too low to up-scatter CMB photons via ICS into the \textit{Fermi}-LAT window (>100 MeV). Therefore the ICS contribution can be safely neglected in the calculation of the PBH constraints. However, the ICS can dominate over the prompt $\gamma$-rays for higher dark matter masses for which the ICS signal >100 MeV, see e.g. the dashed-dotted line of the left panel. This shows that neglecting the ICS signal by only considering prompt $\gamma$-rays introduces a significant error in $\gtrsim$ 500 GeV WIMPs as has been done in other works.

\bibliography{main}

\end{document}